\documentclass[%
reprint,
10pt,
nofootinbib,
amsmath,amssymb,
aps,
]{revtex4-1}

\usepackage{graphicx}
\usepackage{dcolumn}
\usepackage{bm}
\usepackage{lipsum}
\usepackage{xfrac}
\usepackage{bigints}
\usepackage{amsmath,amssymb} 
\usepackage{mathrsfs}
\usepackage{amsfonts}
\usepackage{amsmath,MnSymbol}
\usepackage{pgfplots}
\pgfplotsset{compat=1.5}
\usepackage{floatrow}
\usepackage{float}
\usepackage{soul}
\usepackage{xcolor}
\usepackage[colorlinks=true,linkcolor=blue,citecolor=Green, urlcolor=teal]{hyperref}%
\usepackage{footmisc}
\usepackage{footnote,caption}
 
\def\IB#1{\boldsymbol{#1}} 
\def\W/!i#1{\Wi} 
\def\bten#1{\IB{\mathsf{#1}}}
\def\ang#1{{\langle {#1} \rangle}}  

\definecolor{Green}{HTML}{167425}
\definecolor{azure}{HTML}{005FFF}

\begin{document}
	

	\preprint{APS/123-QED}
	
	\title{Orientational dynamics and rheology of  active suspensions in weakly viscoelastic flows}
	
	\author{Akash Choudhary\textsuperscript{1,3}}
	\homepage{achoudhary@iitk.ac.in}
	\author{Sankalp Nambiar\textsuperscript{2}}
	\author{Holger Stark\textsuperscript{1}}
	\homepage{holger.stark@tu-berlin.de}
	\affiliation{\vspace{2mm}
		\textsuperscript{1}Institute of Theoretical Physics, Technische Universit\"{a}t Berlin, Hardenbergstr. 36, 10623 Berlin, Germany\\
		\textsuperscript{2}Nordita, KTH Royal Institute of Technology and Stockholm University, Stockholm 10691, Sweden\\
		\textsuperscript{3}Department of Chemical Engineering, Indian Institute of Technology Kanpur, Uttar Pradesh 208016, India
	}%

	\begin{abstract}
	\noindent
	\textbf{Abstract.}
	Microswimmer suspensions in Newtonian fluids exhibit unusual macroscale properties, such as a superfluidic behavior, which can be harnessed to perform work at microscopic scales.
	Since most biological fluids are non-Newtonian, here we study the rheology of a microswimmer suspension in a weakly viscoelastic shear flow. At the individual level, we find that the viscoelastic stresses generated by activity substantially modify the Jeffery orbits well-known from Newtonian fluids. The orientational dynamics depends on the swimmer type; especially pushers can resist flow-induced rotation and align at an angle with the flow.  To analyze its impact on bulk rheology, we study a dilute microswimmer suspension in the presence of random tumbling and rotational diffusion. Strikingly, swimmer activity and its elastic response in polymeric fluids alter the orientational distribution and substantially amplify the swimmer-induced viscosity.
	This suggests that pusher suspensions reach the superfluidic regime at lower volume fractions compared to a Newtonian fluid with identical viscosity. 
\end{abstract}
\maketitle

\section*{Introduction}

Systems of particulate matter suspended in fluids are prevalent in numerous natural and industrial processes.
Active suspensions are particulate systems that are driven out of equilibrium by converting chemical energy or fuel into mechanical work to achieve self-propulsion \cite{marchetti2013Review,zottl2016emergent}.
Motile microorganisms are the prototypical example of
active motion that is generated via metachronal actuation of hairlike cilia (\textit{Paramecium, Volvox}), by whipping cell-attached flagellar appendages (spermatozoa,  algae), or by rotating a bundle of helical flagella (bacteria) \cite{lauga2020fluid}.
Microswimmers navigate through their environment by sensing or interacting with gradients in hydrodynamic, chemical, thermal, and light 
fields; a strategy known as taxis \cite{berg1972chemotaxis,demir2012bacterial,stark2018artificial,mathijssen19,jing2020chirality,doan2020trace,ramamonjy2022nonlinear}.
A particular example is rheotaxis, where microswimmers 
experience hydrodynamic gradients and swim against the flow, which
is relevant for biofilm formation and reproduction \cite{suarez2006sperm,zottl2012nonlinear, zottl2016emergent,jing2020chirality}.

Pathogenic microswimmers when infiltrating human and animal bodies have to pass through mucus linings that lubricate 
and protect our respiratory tracks, eyes, urogenital and gastrointestinal systems \cite{sznitman2015locomotion}. 
The presence of mucin fibres (typically 3-10 nm in length \cite{mucus_length_shogren1989}) and DNA  makes these mucus linings viscoelastic and shear thinning \cite{mucus_lai2009micro}. 
The 
linings are often subjected to shearing motion, for instance, during blinking, coughing, reproduction, and continuos mucociliary 
clearance in respiratory systems.
In such microbiological flows, the non-Newtonian fluid properties can alter the swimmer's rheotactic behaviour \cite{mathijssen2016upstream,choudhary2022soft}.




In this article we study the bulk rheology
of microswimmers in viscoelastic flows. For Newtonian fluids, several rheological experiments on suspensions of extensile swimmers like \emph{E.coli} have shown that activity can drastically reduce the effective viscosity \cite{sokolov2009reduction,gachelin2013non,mcdonnell2015motility}, even down to the `superfluidic' limit \cite{lopez2015turning}. 
The mechanism, first outlined by Hatwalne \emph{et al.} \cite{hatwalne2004rheology} and elaborated further by Haines \emph{et al.} \cite{haines2009three} and Saintillan \cite{saintillan2010dilute},
is as follows. The orientations of elongated swimmers in shear flow follow periodic Jeffery orbits \cite{jeffery1922motion} and are also subject to thermal noise. This competition yields a mean orientation that points in the extensional quadrant of applied shear flow. 
With such an orientation
extensile microswimmers (pushers) support the applied shear
flow and thereby reduce the effective shear viscosity, whereas contractile swimmers like \emph{C. reinhardtii} (pullers) resist it and thus increase viscosity \cite{rafai2010effective}.

Since almost all biological fluids are non-Newtonian, there has been a recent interest towards developing theoretical 
frameworks that capture the individual and collective dynamics of microswimmers in complex fluids \cite{li2016collective,li2021microswimming,spagnolie2022swimming}.
For example, experiments have shown reduced tumbling and increased persistence lengths of bacteria in polymeric 
fluids \cite{patteson2015running,zottl2019enhanced,kamdar2022colloidal}.
Viscoelasticity can also initiate spatiotemporal order in active suspensions.
For example, a recent study found that DNA polymers trigger oscillatory vortices in confined suspension of \textit{E.coli} \cite{liu2021viscoelastic}.
However,  determining the effective shear viscosity of an active suspension in a viscoelastic fluid
has been an uncharted territory because of its complexity:  the elastic relaxation of non-Newtonian fluids and their 
shear-thinning/thickening property.

To gain better insights into biological and artificial motility in biological fluids, this communication investigates the role of non-linear polymeric stresses in non-Newtonian fluids. Specifically, this non-linearity allows us to
directly couple a background shear flow to the active disturbance flow generated by a microswimmer. 
For spherical microswimmers in a Poiseuille flow, we already showed that 
due to such a coupling they experience a swimming lift force that
depends on the swimmer type \cite{choudhary2022soft}.
Here, we will show that this nonlinear coupling influences the Jeffery orbit of an elongated swimmer in a shear flow, which thereby also fundamentally alters the bulk rheological response. Such activity-induced changes in the orientational dynamics
cannot be observed in Newtonian fluids because the viscous stress tensor does not permit such a nonlinear coupling. 
De Corato and D'Avino \cite{corato2017dynamics} recently showed that a spherical microswimmer in the shear flow of a non-Newtonian fluid	can exhibit a rotational velocity that is different from a Newtonian fluid.
Although qualitative changes did not occur for weak viscoelasticity,
strong viscoelasticity affected the orientational dynamics of pushers, pullers, 
and neutral swimmers distinctly.
Here we consider an elongated particle and show that even weak viscoelasticity modifies the orientational dynamics,
and ultimately, the rheology of a microswimmer suspension.

The current work uses the model of a second-order fluid with a single elastic relaxation time that captures the dynamics of 
polymeric fluids in the dilute limit (Boger fluids) \cite{bird1987dynamics,james2009boger}. 
For weak elasticity, quantified by the Weissenberg number, we perform a perturbative analysis.
Thereby we show that  
the inherent non-linearity in the polymeric stress tensor together with activity significantly alters the Jeffery orbits known from the Newtonian fluid and orbits of passive rods in a viscoelastic fluid \cite{leal1975slow,brunn1977slow}.
To evaluate the influence of thermal noise, we combine our results with the orientational Smoluchowski equation and find that the altered deterministic dynamics
also substantially affects the orientational distribution, known as the suspension microstructure \cite{hinch1975constitutive}.
It strongly differs between extensile (pushers) and contractile (pullers) microswimmers. The orientational distribution allows
us to directly determine the effective viscosity of the active suspension from an orientational average over the swimmer stresslets.
Our analysis shows that  fluid elasticity reduces the effective viscosity of active suspensions for both pushers and pullers
compared to Newtonian fluids. 
The reduction increases with activity and might allow to reach the superfluidic limit for smaller swimmer densities.

\section*{Results}

\subsection{Setup and second-order fluid}
\begin{figure}[t]
	\centering
	\includegraphics[width=1\textwidth]{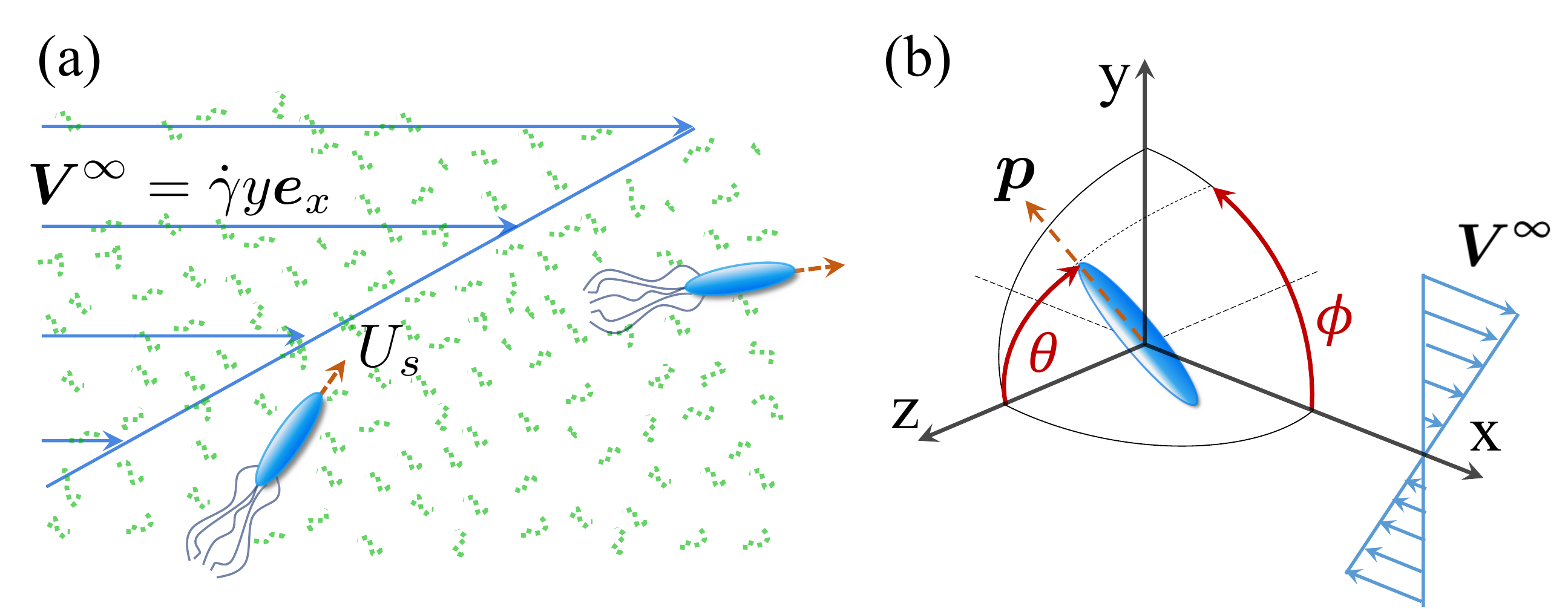}
	\caption{\textbf{Microswimmers in viscoelastic shear flow}. Schematics showing (a) the dilute suspension of microswimmers in an external shear flow $ \IB{V}^{\infty} $ of a viscoelastic fluid, (b) coordinate system 
		moving with an individual swimmer that is modelled as an active prolate spheroid of major axis $ a $ and minor axis $ b $. Here $ \IB{p} $ and $ U_s $ denote the orientation and swimming speed, respectively.}
	\label{schematic}
\end{figure}

\noindent
We consider a dilute suspension of microswimmers in a shear flow of a viscoelastic fluid, for instance, consisting of polymers 
dissolved in a Newtonian fluid as shown in Fig.\ \ref{schematic}(a). Figure\ \ref{schematic}(b) depicts the coordinate system 
moving with the microswimmer that is modelled as an active prolate spheroid and swims with speed $ U_s $. 
The uniformly distributed polymers are much smaller than the microswimmers and hence modeled within a continuum description. 
The inertia-less hydrodynamics is governed by the mass and momentum conservation as $ {\IB{\nabla}} \cdot \IB{V} = 0$ and $ {\IB{\nabla}} \cdot \bten{T}=0  $, respectively, where $ \IB{V} $ is the  velocity field and $ \bten{T} $ is the total stress tensor.
It follows the second-order fluid (SOF) model: 
$ \bten{T} =-P \, \bten{I} + 2 \, \bten{E} + \text{Wi} \, \bten{S} $ \cite{bird1987dynamics}.
Here, $ \bten{E} $ denotes the rate-of-strain tensor and
$
\bten{S} =	  4 \bten{E}\cdot \bten{E} + 2 \delta \overset{\Delta}{\bten{E}}
$
is the polymeric stress tensor, which is quadratic in $ \bten{E} $ and contains the lower-convected time derivative.
The SOF model not only allows us to capture the elastic effects pertinent to Boger fluids, but also to obtain analytical results  for small $ \text{Wi} $.
Since we consider a steady shear rate in this work, we disregard the partial time derivative of $\bten{E}$.
In above governing equations, length, velocity, and pressure are  already non-dimensionalized by the swimmer length ($ l=2a $), $ \dot{\gamma} l$, and $ \mu_f \dot{\gamma} $, respectively, where $ \mu_f $ is the fluid shear viscosity. 
Also, the stress tensor is written in dimensionless units using characteristic numbers (\text{Wi} and $ \delta $).
In particular, $ \text{Wi} =  t_{\text{relax}} \dot{\gamma} $ is the shear based Weissenberg number that quantifies the importance of elasticity in the medium.
It compares the shear rate $ \dot{\gamma} $ or inverse shearing time to the polymer relaxation time $ t_{\text{relax}} = (\Psi_{1}+\Psi_{2})/\mu_f $, where $ \Psi_{1}$ and $\Psi_{2} $ are the normal stress coefficients.  
Furthermore, $ \delta = -\Psi_{1}/2(\Psi_{1}+\Psi_{2}) $ is the viscometric parameter that typically varies between $ -0.7 $ to $ -0.5 $. 
In `Methods: Hydrodynamic model' we explain in detail how we solve the governing equations using a  systematic perturbation expansion in $ \rm{Wi} $ in the limit of weak viscoelasticity.

\subsection{Active spheroid in viscoelastic shear flow}
\noindent
To determine the dynamics of an active spheroid, we first note that a swimmer disturbs the flow field both passively (due to its rigid body) and actively (due to self-propulsion). 
The disturbance fields are implemented using hydrodynamic multipoles
\cite{chwang1975hydromechanics,spagnolie_lauga_2012}.
Flagellated microswimmers like \emph{E.coli} and \emph{Chlamydomonas} generate a force-dipole flow field and 
higher order disturbances: source dipole, rotlet dipole, and force quadrupole.
We included all four of them and found that only the force-dipole flow field affects the swimmer dynamics in leading order in $ \text{Wi} $. 
The force-dipole field around the active spheroid is $ \sigma \frac{\hat{\IB{r}}}{r^2} [3 (\hat{\IB{r}} \cdot \IB{p})^2 -1]$ 
with $\hat{\IB{r}} = \IB{r}/r$.
%
Here, $\sigma$ is a non-dimensional parameter equal to the ratio of the force-dipole strength ($\sigma^*$)
to the stresslet imposed by the shear flow 
($ 8\pi \mu_f \dot{\gamma} l^{3} $), where $\sigma>0$ represents pushers 
and  $ \sigma<0 $ pullers. 
For the current work, we focus on swimmers of $ 5 \mu $m size and shear rates of order $ 0.5-5\text{s}^{-1}$, which corresponds to $ \sigma $ of typical wild-type \emph{E.coli} being roughly $ 0.04-0.1 $ \cite{berke2008hydrodynamic,drescher2010direct}.
Here, the Weissenberg number can be tuned either by varying the shear rate or relaxation time of the fluid. The latter for the current system is $ t_{\text{relax}} \lesssim 0.1s $, which for instance, can be realised in PEO solutions of molecular weight ranging from 2 $ - $ 4 $ \times 10^{6} $ g mol\textsuperscript{-1} and concentration between 0.25 $ - $ 0.5 wt$ \% $ \cite{ebagninin2009rheological}. We vary $ \text{Wi} $ from $ 0.05 -0.5 $ by fixing $ t_\text{relax}$ and changing the shear rates between $ \dot{\gamma} \sim 0.5 - 5 \text{s}^{-1}$, following a protocol similar to that for varying $\sigma$.

\begin{figure}[t]
	\centering
	\includegraphics[width=1\textwidth]{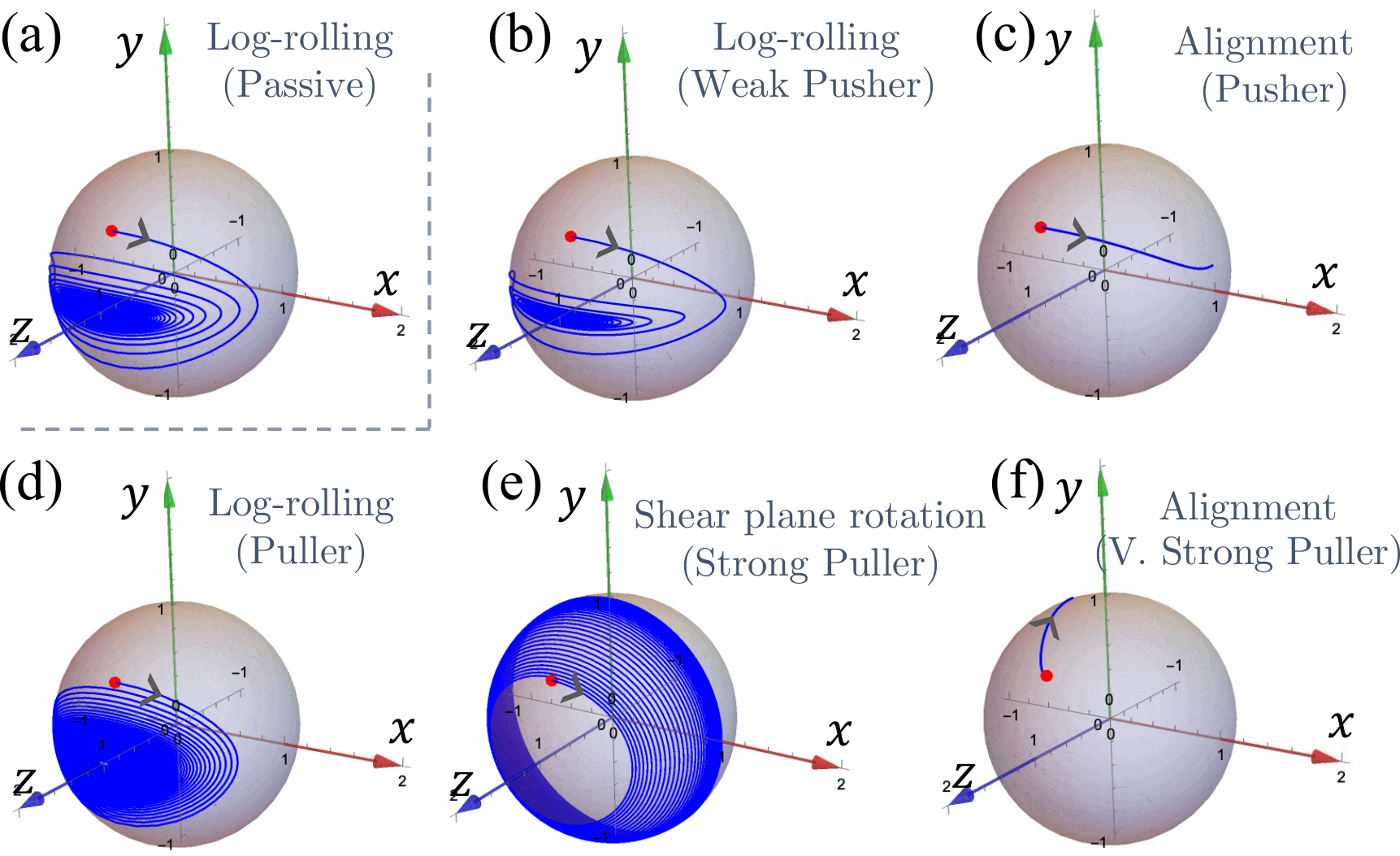}
	\caption{\textbf{Orientation dynamics} of passive $(\sigma =0)$	and active particles in weakly viscoelastic shear flow. The blue curves show time traces of the orientation vector $\IB{p}$ on the unit sphere starting at the red dot.
		(a) Dipole strength $\sigma = 0 $, (b) $ \sigma=0.06 $, (c) $ \sigma=0.2 $, (d) $ \sigma=-0.2 $, (e) $ \sigma=-0.8 $, (f) $ \sigma =-2.4 $ . Other 
		parameters: Weissenberg number $ (\text{Wi})=0.2$, aspect ratio $(\lambda)=5$, initial polar angle $(\theta_{0})=\pi/5$, initial azimuthal angle $(\phi_{0})=\pi/2$, viscometric parameter $ \delta=-0.6 $. 
		The active and passive viscoelastic coefficients of Eq. (\ref{EOM}) are evaluated to be $\alpha_{1}=4.62$, $\alpha_{2}=-0.33$, 		$\beta_{1}=0.68$, and $\beta_{2}=1.46 $.}
	\label{fig:orbits}
\end{figure}

In Newtonian fluids, the equation of motion for the orientation $ \IB{p} \, (\theta, \phi)$ gives the Jeffery orbits. 
To formulate this equation for viscoelastic shear, we note that  the polymeric stress tensor $ \bten{S} $ is quadratic in the rate-of-strain tensor and vorticity \cite{bird1987dynamics}. 
Thus, similar to Einarsson \emph{et al.}\ \cite{einarsson2015rotation} we determine all terms that by symmetry contribute to the rate of change $ \dot{\IB{p}} $ up to first order in $ \text{Wi} $. Neglecting small terms, we arrive at 
\begin{align}\label{EOM}
	\dot{\IB{p}} &=   \left(  \bten{I} -  \IB{pp}  \right) \cdot  \left(   \bten{E}^{\infty} \cdot \IB{p}   \right) \left[  \Lambda + \text{Wi}  \, \sigma  \alpha_{1} + \text{Wi} \, \beta_1 \,  \bten{E}^{\infty}: \IB{pp} \right] 
	\nonumber \\
	& \; + \IB{\Omega}^{\infty} \times \IB{p} \left[  1 + \text{Wi} \, \sigma  \alpha_{2} + \text{Wi} \, \beta_2 \,  \bten{E}^{\infty}: \IB{pp}   \right] + O(\text{Wi}^{2})
\end{align}
in non-dimensional form, where $ \IB{\Omega} $ is the angular velocity and the superscript $ \infty $ denotes the quantities that belong to the prescribed 
shear flow.
%
The shape factor $ \Lambda = \frac{-1+\lambda^{2}}{1+\lambda^{2}}$ contains the aspect ratio
$ \lambda = a/b$,
the ratio of major to minor axis.
The shape factor approaches  $ +1 $ and $ -1 $ for needle and disk-like particles, respectively. 
Since microswimmers are usually elongated, we focus on prolate spheroids of $ \Lambda > 0.9 $, which corresponds to $\lambda > 4$; 
it also helps in simplifying the calculations (see `Methods: Hydrodynamic model').
The terms with coefficients $ \alpha_i$ and $ \beta_i $ represent the active and passive viscoelastic contributions, respectively. 
These coefficients are evaluated explicitly as outlined in `Methods: Orientation dynamics' using the Lorentz reciprocal theorem, where we also derive Eq.\ (\ref{EOM}).
For our relevant parameters, we find $\alpha_1 \gg |\alpha_2|$. 
This along with Eq.\ (\ref{EOM})
indicates that the modification arising from the activity predominantly depends on the extensional part  ($ \bten{E}^{\infty} $) of the viscoelastic shear flow   rather than its rotational component.
Since the coefficients vary only weakly with $\lambda$ and $ \delta $, they are treated as constants in the following discussion of results (see also Supplementary Note 2A).

\begin{figure}[t]
	\centering
	\includegraphics[width=1\textwidth]{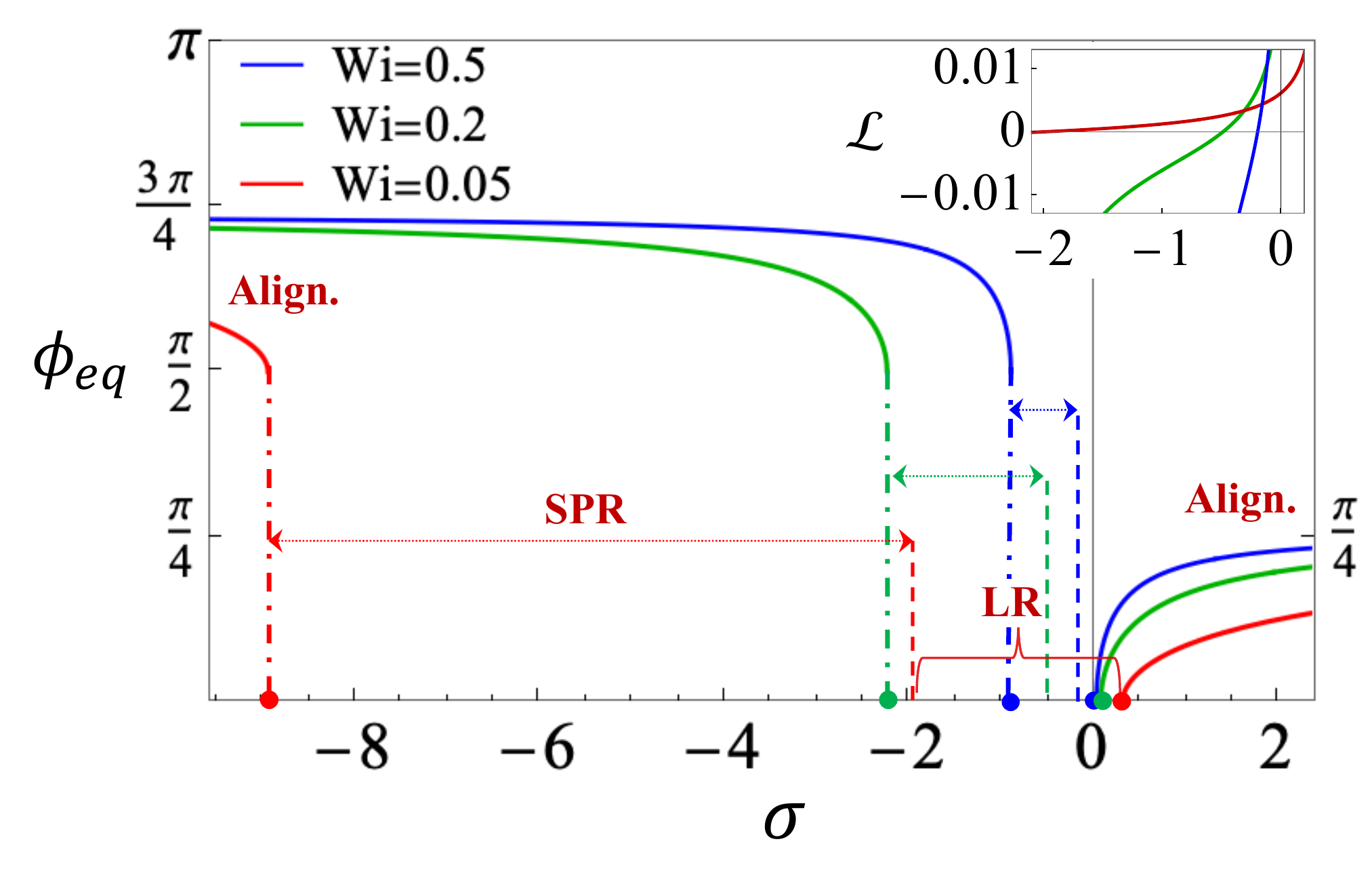}
	\caption{\textbf{State diagram} showing the different states of the orientational dynamics for three Weissenberg numbers. 
		With increasing dipole strength $\sigma$ from left to right, we observe  alignment (Align.), shear-plane rotation (SPR), 
		log rolling (LR) and again alignment.
		In the alignment state, the alignment angle $\phi_{eq}$ is plotted versus dimensionless activity $\sigma$  as solid lines.
		The dots on the horizontal axis indicate the critical values $\sigma_c$ at which alignment (Align.) occurs 	either to the left of the dashed-dotted lines (pullers, $\sigma < 0$) or to the right (pushers, $\sigma>0$). The dashed lines separate SPR and LR from each other at $\sigma = -1.93, \, -0.49, \, -0.19 $ for increasing $ \text{Wi}$.
		The inset shows the variation of Lyapunov exponent ($ \mathcal{L} $) with $\sigma$.
		Other parameters are chosen as in Fig.\ \ref{fig:orbits}.
	}
	\label{fig:phase}
\end{figure}

In the absence of polymers ($ \text{Wi} = 0 $), Eq.\ (\ref{EOM}) reduces to the well-known Jeffery equation \cite{jeffery1922motion} that has an infinite number of neutrally stable solutions, \emph{i.e.,} the microswimmer's orientation traces periodic orbits that depend on initial conditions. During orbital time period $ T = 2\pi (\lambda+\lambda^{-1})/\dot{\gamma} $, they spent the majority of their time  ($ \propto \lambda $) aligned near the flow-vorticity plane.
When polymers are present, the passive viscoelastic effect breaks the degeneracy of Jeffery orbits and the microswimmer 
slowly drifts towards an alignment with the vorticity axis, known as the ``log-rolling'' state \cite{gauthier1971_shear-flow,brunn1977slow,davino2015particle}. 
Figure \ref{fig:orbits}a shows the curve traced by the orientation of a spheroid on a unit sphere while drifting towards the vorticity axis.
This dynamics can be deduced from Eq. (\ref{EOM}) for passive spheroids ($\sigma = 0$) when the terms with coefficients 
$ \beta_{1} $ and $ \beta_{2} $ are present.
%
%
In Supplementary Note 2C  we also compare our results of passive spheroids with Brunn \cite{brunn1977slow}.


Now, we illustrate the influence of activity.
For weak pushers ($0 < \sigma \lesssim 0.1 $), the orbits towards log rolling look similar to the ones of passive spheroids,
albeit with a larger aspect ratio (see Fig.\ \ref{fig:orbits}b) since they exhibit more skewed trajectories.
This can directly be
inferred from Eq.\ (\ref{EOM}),
where  activity ($\sigma$) in the term with $\mathrm{Wi} \, \sigma \alpha_1$  modifies the shape factor.
Since $ \alpha_{1}>0 $, a pusher ($ \sigma > 0 $) effectively increases $\Lambda$, which corresponds to a higher aspect ratio. 
The opposite occurs for weak pullers ($ \sigma < 0 $) as they  behave like a passive spheroid with reduced aspect ratio.
Here the trajectories are more circular as shown in Fig. \ref{fig:orbits}d.

As the dipole strength of a pusher increases beyond a critical value $ \sigma_{c} $, the orientation drifts to the shear plane ($ \theta=\pi/2 $) and aligns at an angle $ \phi_{eq} $ with the flow direction (see Fig.\ \ref{fig:orbits}c).  
We quantify this transition in Fig.\ \ref{fig:phase} and show  that $ \sigma_{c} $ (dots close to 0) decreases with increasing ${\rm Wi}$ while $ \phi_{eq} $ increases with both $\sigma$ and $ \rm{Wi} $. 
This dynamical behaviour can again be discerned from Eq.\ (\ref{EOM}), which essentially balances the effect of rotational ($\IB{\Omega}^{\infty}$) and elongational ($\bten{E}^{\infty} $) flow on $\IB{p}$.
Now, activity together with viscoelasticity allows to control this balance. In particular, for our case of $\alpha_1 \gg |\alpha_2|$, we expect the elongational flow to dominate the dynamics for increasing activity. Indeed, when the effective shape factor $\Lambda + \text{Wi}  \, \sigma  \alpha_{1}$ exceeds unity at a critical value $\sigma_c$, the dynamics of $\IB{p}$ transitions from the orbital to the alignment state as favored by the elongational flow.  
In `Methods: Dynamical analysis 1', we show for a  modified dynamical system that the relevant eigenvalue of the dynamical matrix becomes real at $\sigma_c$, as expected for 
such a transition, and the corresponding eigenvector gives the alignment angle $\phi_{eq}$. 
	As $\sigma$ increases, 
	$\phi_{eq}$ grows from zero and approaches $\pi/4$ for large $\sigma$, since in this limit, the elongational flow ($\bten{E}^{\infty}$) with its principal axis along $\phi=\pi/4$ completely dominates the dynamics of $\IB{p}$.

	For pullers, Fig.\ \ref{fig:orbits}e shows that increasing the activity induces a transition from log rolling towards rotation in the shear plane.
	This orbit is also observed for passive particles with oblate shape ($ \Lambda<0 $) in viscoelastic flows \citep{gauthier1971_shear-flow,brunn1977slow,davino2015particle}.
	It appears here and acts as an attracting limit cycle, when the effective shape factor $\Lambda + \text{Wi} \sigma \alpha_1$ in Eq.\ (\ref{EOM}) 
	becomes negative due to the puller activity $\sigma <0$.
	%
	We have performed a stability analysis for the shear-plane rotation in `Methods: Dynamical analysis 2' and calculated the Lyapunov stability exponent $ \mathcal{L} $. 
	It determines the exponential time variation of the disturbed limit cycle. 
	In the inset of Fig.\ \ref{fig:phase}, we show the Lyapunov exponent as a function of $\sigma$.
	For weak and moderate pullers, $ \mathcal{L} >0$ shows that shear-plane rotation is an unstable limit cycle, where the system drifts towards stable log rolling.
	As the dipole strength of the puller becomes more negative, $ \mathcal{L}$ 
	turns negative meaning that shear-plane rotation is stable. 
	Upon analysing the orbital dynamics in the shear-plane rotation, we find that increasing $ |\sigma| $ gradually slows down the 
	swimmer's rotation near the $y$-axis, along which the flow gradient is applied.
	Eventually, a transition to permanent alignment with $\phi_{eq} = \pi/2 $ occurs, which can be similarly analyzed as the alignment transition of pushers. However, here it occurs at the effective shape factor $\Lambda + \text{Wi}  \, \sigma_c  \alpha_{1} = -1$. With further increasing $|\sigma|$ the 
	active spheroid tilts against the flow, in contrast to pushers, and asymptotes at $ 3\pi/4 $, which is the direction of the second principal
	axis of the elongational flow ($\bten{E}^{\infty}$).
	This behavior is illustrated in Fig.\ \ref{fig:orbits}f and Fig.\ \ref{fig:phase}. Finally,  Fig.\ \ref{fig:phase} also shows that as
	$ \text{Wi} $ increases, the regime of shear-plane rotation shrinks.

	\subsection{Impact of noise on orientational dynamics}
	\noindent
	The deterministic orientational dynamics 
	is disturbed by two types of stochastic reorientations of the  swimming direction, which we now address with the help of the Smoluchowski equation. 
	First, a bacterium tumbles, which is triggered when the rotation of one of its flagella reverses so that it leaves the  flagellar bundle
	\cite{berg1983random,lauga2016bacterial,adhyapak2016dynamics}. 
	For a wild type \emph{E.coli}, tumbling occurs roughly every $1\mathrm{s} $, where it attains a new random orientation.
	Although tumbling is biased in the forward direction with a mean tumbling angle of roughly $68.5^\circ$, we model it to be unbiased for computational ease because it only affects our results marginally as suggested by Nambiar \emph{et al.}\ \cite{nambiar2017stress}.
	Second, thermal rotational diffusion continuously reorients a microswimmer but its effect is small compared to tumbling.

	We now evaluate the orientational distribution of an ensemble of non-interacting microswimmers in steady state, caused by the three mechanisms discussed so far:  deterministic motion in background shear flow, tumbling, and rotational diffusion.
	The steady-state probability distribution function $ \psi(\IB{p}) $ for the orientation vector of non-interacting microswimmers
	is governed by the Smoluchowski equation \cite{doi1988theory}
	\begin{equation}\label{Smol}
		\text{Pe}_f \nabla_{p} \cdot \left( \IB{\dot{p}} {\psi} \right)    - \tau D_r  \nabla_{p}^{2} {\psi} +  \left( {\psi} -  \frac{1}{{4\pi}} \right) = 0,
	\end{equation}
	where $\psi(\IB{p})$ is normalized to unity.
	The first term describes the orientational drift using $\nabla_{p} $ as the gradient operator on the unit sphere. To compare the 
	strength of flow-induced reorientation to the mean time $\tau$ between two tumbling events,
	we introduce the flow P\'eclet number $\mathrm{Pe}_f = \dot{\gamma} \tau$. Tumbling away from the swimming
	direction $\IB{p}$ is handled by the third term and rotational diffusion by the second term. 
	For bacteria, the latter is typically  small compared to tumbling since $ \tau \sim 1 \mathrm{s} $ \cite{berg1983random} and $ D_r  \lesssim 0.1 \mathrm{s}^{-1} $, as an estimate from the Stokes-Einstein relation shows.
	In restricting ourselves to Eq.\ (\ref{Smol}), we assume a spatially uniform system valid when hydrodynamic and steric
	interactions with bounding surfaces can be neglected \cite{spagnolie_lauga_2012,spagnolie2022swimming}. 
	In particular, this means that the mean length of persistent swimming, $U_s \tau$, where $U_s$ is the swimming speed, is much
	smaller than the spatial extent of the system. 
	Finally, we consider weak fluid elasticity with relaxation time $t_{\mathrm{relax}} \ll \tau, D_r^{-1}$, 
	so that the fluid relaxes faster than the time scales given by stochastic reorientations
	\cite{cohen1987orientation}.
	Typical dilute and semi-dilute PEO solutions of molecular weight of $ \sim 10^{6}$ g mol\textsuperscript{-1}  have $ t_{\text{relax}}  \lesssim 0.1 $s \cite{ebagninin2009rheological}, whereas the stochastic reorientation time scale is always $ 1 $s or larger.
	Thus, here we do not need to take into account any memory in the rotational noise.
	
	In the following, 
	we want to emulate a rheological experiment and vary the shear rate $ \dot{\gamma} $ via $ \text{Pe}_f $,
	while the fluid and swimmer properties are kept constant. Therefore, in Eq.\ (\ref{EOM}) for the rotational 
	drift velocity $\dot{\IB{p}}$, we rewrite $\text{Wi} $ as $ \text{Pe}_{f} \text{De} $, where the Deborah number 
	$ \text{De} = t_{\mathrm{relax}}/\tau$ compares the fluid relaxation time to $\tau$.
	Furthermore, the second relevant parameter, $ \text{Wi} \sigma $, does not 
	explicitly depend on $\dot \gamma$ but rather quantifies the strength of activity relative to fluid elasticity. 
	To vary the activity  independent of other parameters, we replace $ \text{Wi} \sigma $ by
	$ \text{Pe}_{a} \text{De}/8\pi$. Here, $ {\rm Pe}_a $ is the signed activity Peclet number that is positive for pushers and negative for pullers. 
	Using $U_s \sim \sigma^{*} / (\mu_f  l^{2})$ for the swimming speed with the force dipole moment $\sigma^{*}$\cite{berke2008hydrodynamic}, we can write it in the familiar form $ {\rm Pe}_a \sim U_s \tau/l $. 
	Thus, its magnitude compares the persistence length to the body length $l$. A wild-type \emph{E.coli} typically has 
	$ \text{Pe}_{a} \lesssim 5 $ \cite{berg1983random,chattopadhyay2006swimming}. 
	We numerically solve Eq.\ (\ref{Smol}) for arbitrary $Pe_f$ by expanding $\psi$ in 
	spherical harmonics and 
	taking into account the first 100 harmonics. The numerical solution is also verified analytically in the limit of 
	$ \text{Pe}_f \ll 1 $ (see further details in `Methods: Kinetic model').

\begin{figure}[!t]
		\centering
		\includegraphics[width=0.9\textwidth]{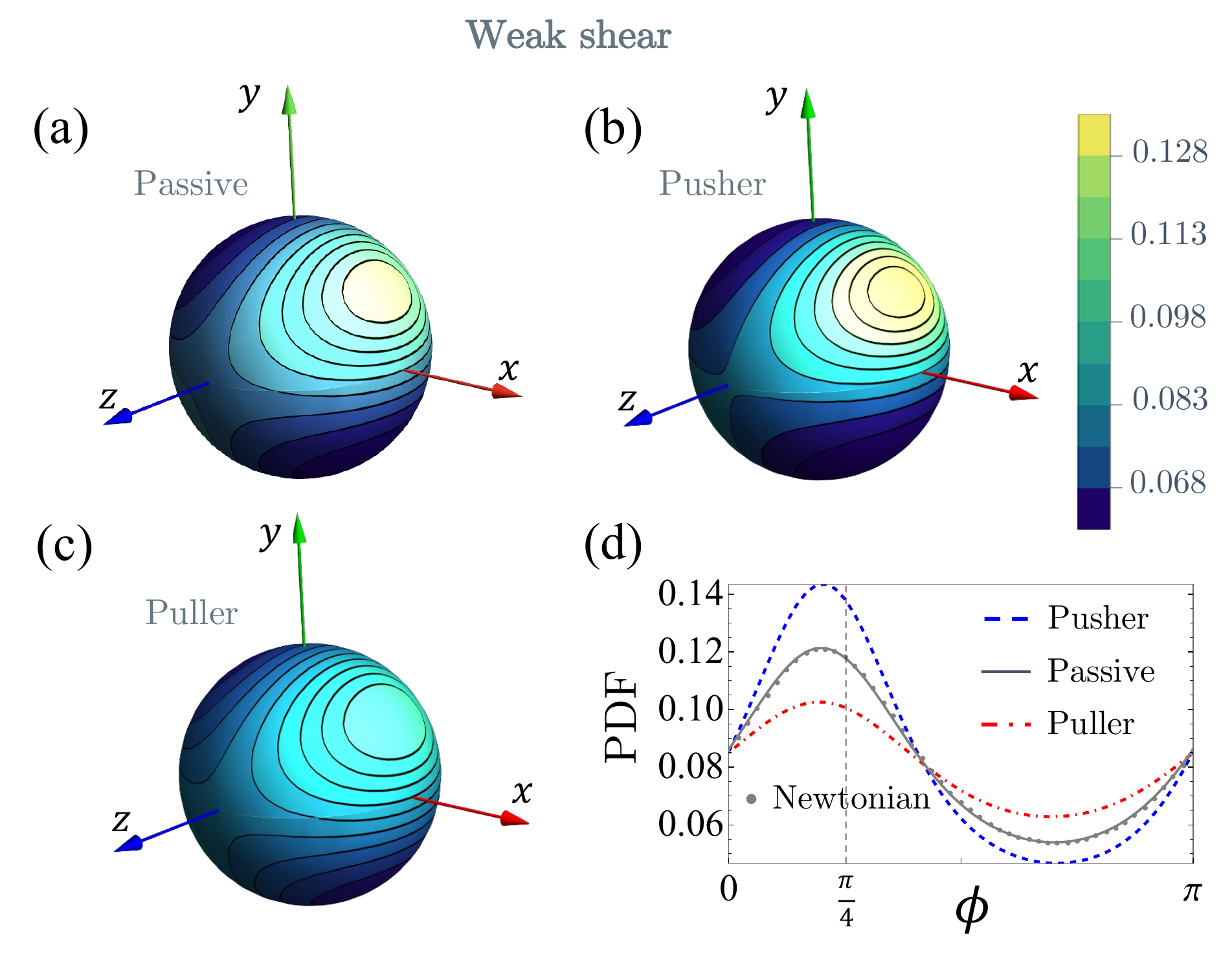}
		\includegraphics[width=0.9\textwidth]{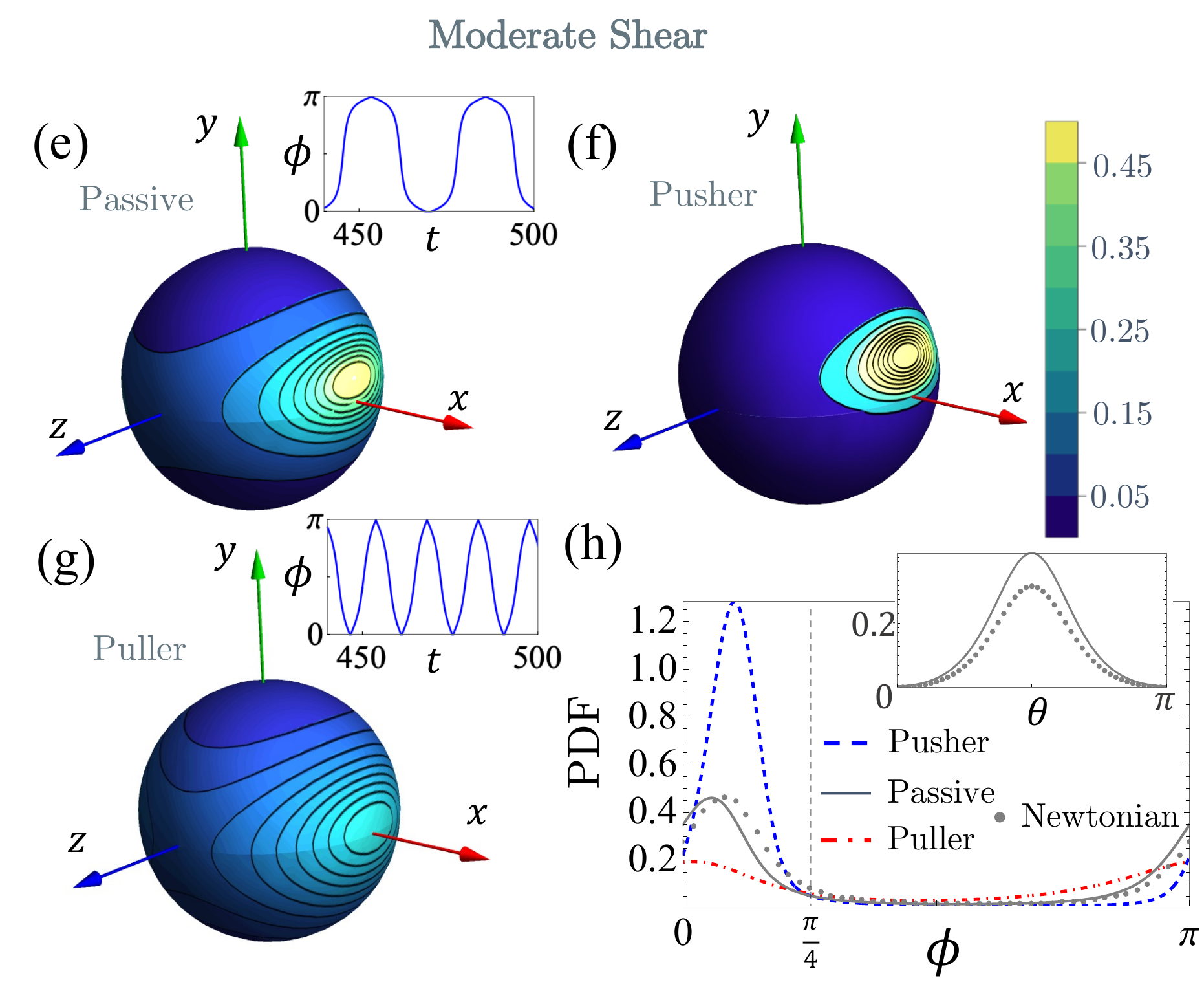}
		\includegraphics[width=0.88\textwidth]{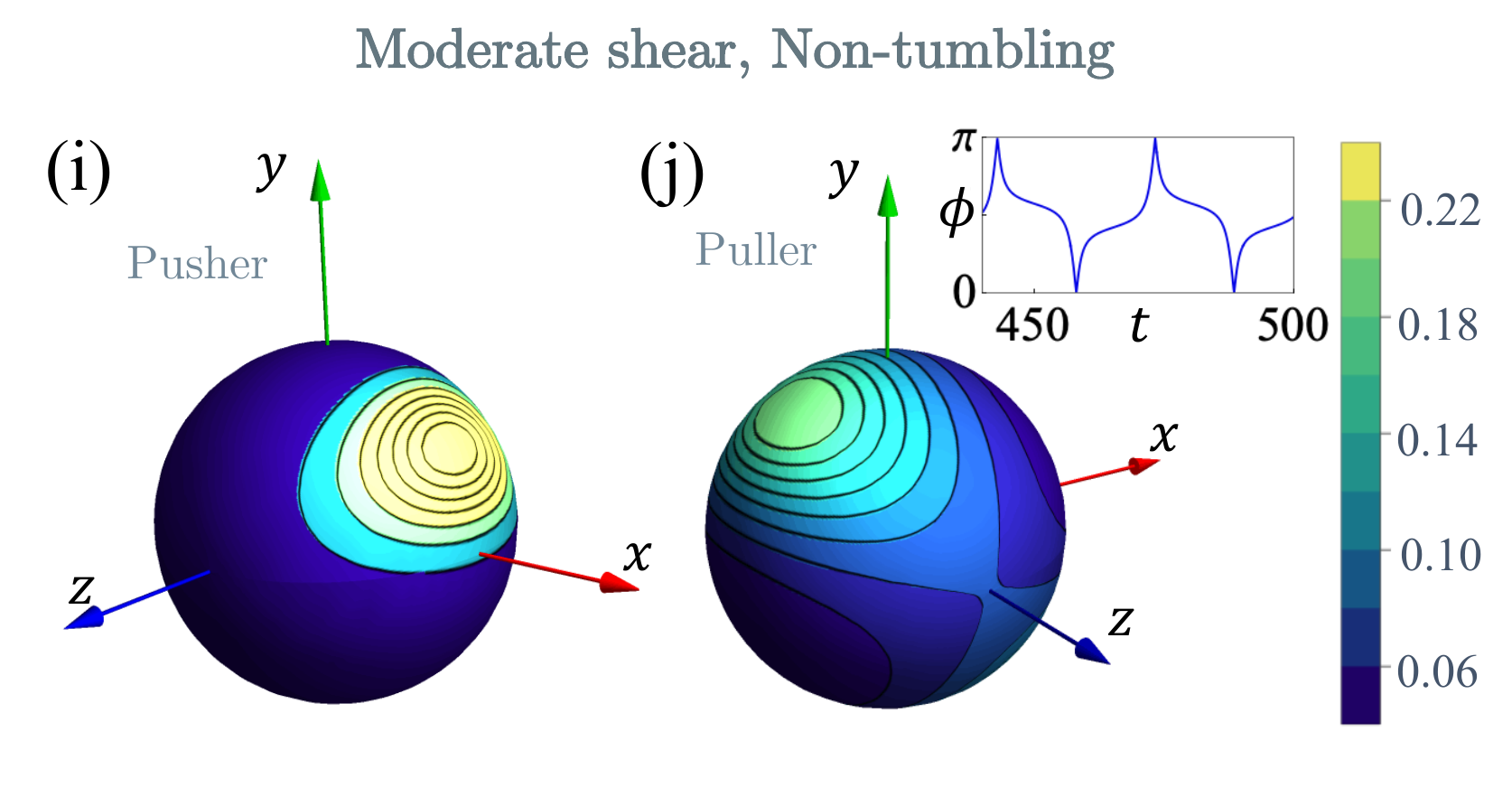}
		\caption{
			\textbf{Orientational probability distribution $\psi(\IB{p})$} for
			(a-d) weak shear flow at flow Péclet number $ \text{Pe}_f = 0.5 $, (e-j) moderate shear flow at $ \text{Pe}_f = 5 $. The activity Péclet number ($ \text{Pe}_a $) varies as (a,e) $ \text{Pe}_a=0 $, (b,f) $ \text{Pe}_a=20 $, (c,g) $ \text{Pe}_a=-20 $ and always with $ D_r \tau =0.1 $, where $ \tau $ is tumbling time and $ D_r $ is the rotary diffusion coefficient.
			(d,h) Probability distribution function (PDF) in the flow-shear plane; the inset in (h) shows PDF in the flow-vorticity plane.
			Non-tumbling particles: (i) $ \text{Pe}_{a}=100 $, (j) $ \text{Pe}_{a}=-100 $. The insets in (e,g,j) show the polar angle $\phi(t)$ for the shear-plane rotation state corresponding to the respective distributions.
			Colour bars indicate the value of PDF.
			Other parameters are chosen as in Fig.\ \ref{fig:orbits} with Deborah number $ \text{De}={t_{\text{relax}} / \tau}=0.1 $.
			To locate these distributions in Fig.\  \ref{fig:phase}, we give the corresponding Weissenberg and activity numbers:	
			(a-d) $ \text{Wi}=0.05 $, $ \sigma=\pm 1.6 $; (e-h) $ \text{Wi}=0.5 $, $ \sigma=\pm0.16 $; (i,j) $ \text{Wi}=0.5 $, 
			$ \sigma=\pm0.8 $.
		}
		\label{fig: microstructure}
	\end{figure}

	Figure \ref{fig: microstructure} shows the orientational probability distribution of passive and active particles
	for various cases.
	It quantifies the orientational `microstructure'  of a suspension of non-interacting and orientable particles subject
	to shear flow
	\cite{hinch1976constitutive,chen1996rheology,nambiar2017stress}.
	We begin with discussing the case of weak shear rate ($\text{Pe}_f  < 1$) in Figs.\ \ref{fig: microstructure}a-d.
	In the absence of shear flow, the microstructure is solely governed by rotational noise and
	is therefore isotropic.
	For weak shear rates the extensional part ($ \bten{E}^{\infty} $) of the shear flow primarily distorts the microstructure,
	which peaks near the extensional axis as noted for Newtonian fluids by Hinch and Leal \cite{hinch1972effect}. 
	We quantify this in `Methods: Kinetic model 1' by solving Eq.\ (\ref{Smol}) via a perturbation expansion in $\text{Pe}_f$ and obtain 
	the first-order correction:
	\begin{equation}\label{pert_Smol}
		\psi_{(1)}	=  \frac{3 }{4 \pi} (\Lambda+\alpha_{1} \text{De} \text{Pe}_{a}/8\pi)      
		\frac{   \bten{E}^{\infty}: \IB{p}\IB{p} }{1+ 6\tau D_r}  
	\end{equation}
	In the absence of activity, there is no deviation from the Newtonian microstructure at this order. 

	In Figs.\ \ref{fig: microstructure}a-d we use a weak shear rate of $\text{Pe}_{f}=0.5$ 
	and $ |\text{Pe}_{a}| = 20$,
	which is
	an activity close to mutated strains of tumbling \textit{E.\ coli} (pusher) \cite{wolfe1988acetyladenylate}
	and the algae \textit{C.\ reinhardtii} (puller).
	The observed distributions follow the trend of Eq.\ (\ref{pert_Smol}). While pushers  ($ \text{Pe}_a >0 $) enhance the alignment along the extensional axis of the applied shear flow, pullers ($ \text{Pe}_a < 0 $) weaken it. 
	%
	We note that the peak of the orientational distributions in the shear plane (plot d) exhibits a small deviation from $\phi=\pi/4$. This is due to the rotational flow 
	($\IB{\Omega}^{\infty}$), which contributes to $\psi(\IB{p})$ in second order in $ \text{Pe}_f$ \cite{chen1996rheology}.

	As the shear rate increases, the deterministic dynamics becomes more visible. We illustrate this in Figs.\ \ref{fig: microstructure}e-h for  $\text{Pe}_f=5 $. 
	Compared to Fig.\ \ref{fig: microstructure}a, the peak in $\psi(\IB{p})$ for passive particles  (Fig.\ \ref{fig: microstructure}e)
	shifts towards the flow axis and becomes more anisotropic along the flow-vorticity plane, as the deterministic velocity $\dot{\IB{p}}$ is smallest in this plane.
	Eventually, $\dot{\IB{p}}$ approaches zero in the deterministic log-rolling state observed in viscoelastic fluids (Fig.\ \ref{fig:orbits}a).
	However, the deterministic log-rolling state is not completely reflected in the distribution because the slow relaxation towards the vorticity axis is always interrupted by the dominating stochastic reorientations
	(see Supplementary note 4, where we illustrate the orientation distribution for a larger weakly Brownian particle that resembles log-rolling).
	Note that due to the fluid elasticity, the peak of $\psi(\IB{p})$ in Fig.\ \ref{fig: microstructure}h is slightly closer to the flow axis compared to the Newtonian case.
	Furthermore, the inset therein also illustrates a broader distribution in the flow-vorticity plane ($\phi = 0 $). 
	Further increasing $ \text{Pe}_{f}$ spreads the distribution even more in the flow-vorticity plane and, ultimately, for $ \text{Pe}_{f} \gtrsim 100 $ the peak develops along the vorticity axis, which is entirely reminiscent of log rolling.
	However, this is a regime which we cannot strictly capture  as our analysis requires $\text{De}$, $\text{Wi} < 1$, which means $ \text{Pe}_{f} < 10$.
	These findings are qualitatively consistent with earlier studies on passive fibre suspensions at moderate shear rates \citep{cohen1987orientation,chung1987orientation} (further illustrated in Supplementary Note 4). 
	
	Figure 4f demonstrates that the activity of pushers in conjunction with the higher shear rate {($ \text{Pe}_{f}=5$)} makes the distribution more anisotropic. According to Fig.4h, {pushers} strongly focus at an angle $\phi=18^\circ$ in comparison to a weaker peak of passive particles near to flow axis. 
	This angle is close to the alignment angle $\phi_{eq}=21^\circ$ of the deterministic state with the parameters $ \text{Wi}=0.5$, $\sigma=0.16$ in Fig.\ \ref{fig:phase}. 
	Conversely to pushers, pullers reduce the anisotropy in the orientational distribution as shown in Fig.4g. 
	We can directly relate this observation to the deterministic velocity $\dot{\IB{p}}$ with parameters $ \text{Wi}=0.5$, $\sigma=-0.16$ that indicate again the log-rolling state; the resulting orbit is similar to that depicted in Fig.\ \ref{fig:orbits}d. 
	Compared to the passive case, the orbit is more circular, 
	which we already related to a reduced effective aspect ratio when discussing the deterministic orientational dynamics.
	Therefore, the orientation is less constrained to the flow-vorticity plane. {This is explained by  the dynamics of the azimuthal angle $\phi$ in the inset of  Fig.\ \ref{fig: microstructure}g that shows lesser time spent in the plane compared to the inset in Fig.\ \ref{fig: microstructure}e}. As a result, the distribution is broader and less anisotropic \citep{leal1971effect}.
	Consequently, the particle is less susceptible to the effect of stochastic reorientations and thus, its distribution has a peak closer 
	to the flow axis.
	Hence, unlike active Newtonian suspensions \cite{saintillan2010dilute}, the microstructure of microswimmers in a viscoelastic fluid is clearly sensitive to their hydrodynamic signature being either pushers or pullers.
	
	We also examined the orientational distributions of microswimmers that do not tumble and only experience rotational thermal noise. 
	Therefore, in this case we eliminate  the last term on the right-hand side of Eq.\ (\ref{Smol}) and redefine our parameters
	$ \text{Pe}_{f} $, $ \text{Pe}_{a} $, and $ \text{De}$ replacing $ \tau $ by $ D_{r}^{-1} $. 
	For passive  particles and activity $ \text{Pe}_a \sim O(10) $, the microstructure is qualitatively similar to the distributions in Figs.\ \ref{fig: microstructure}a-h.
	However, microswimmers that do not tumble can have larger persistence lengths \cite{elgeti2015physics}. 
	In particular, one can genetically modify \emph{E.coli} such that tumbling does not occur \cite{berke2008hydrodynamic,junot2019swimming,wolfe1988acetyladenylate}.
	Therefore, in Figs.\ \ref{fig: microstructure}i and j we show the orientational distributions for $\text{Pe}_{a} = \pm 100$.
	Compared to Fig.\ \ref{fig: microstructure}f, increasing the activity of a pusher moves the peak in the distribution function
	(at $ \phi=34^{\circ} $) even closer to the alignment angle ($ \phi_{\text{eq}}=36^{\circ} $) of the deterministic alignment state.
	For pullers we observe an even more drastic change upon increasing the activity. 
	The peak in the orientational distribution shifts to the quadrant where the compressional part of the shear flow occurs (see Fig. \ref{fig: microstructure}j).
	Interestingly, this is not due to the deterministic alignment of pullers as the parameters ($ \text{Wi}=0.5, \, \sigma=-0.8 $) belong to the shear-plane rotation state (see Fig. \ref{fig:phase}). 
	As already explained in the discussion of Fig.\ \ref{fig:phase}, the rotational velocity $\dot{\IB{p}}$ slows down near $ \phi=\pi/2 $ 
	when the effective shape  factor $\Lambda + \text{Wi}  \, \sigma_c  \alpha_{1}$ becomes negative. Therefore, the active particle resembles a passive oblate spheroid \cite{brown2000orientational}.  
	Indeed, the inset of Fig.\ \ref{fig: microstructure}j shows how the puller orientation spends more time near $ \phi=\pi/2$. This again shows that even in the presence of significant noise the  microstructure is determined by the deterministic dynamics.
	%
	%

	
	\subsection{
		Shear viscosity of active suspensions}
	
	\noindent
	Finally, we evaluate the effective shear viscosity of a dilute active suspension from the total stress tensor,
	$ \bten{\Sigma} = \bten{\Sigma}_{f} + \bten{\Sigma}_{p} $, where subscripts $f$ and $p$ refer to
	fluid and particle contributions, respectively. 
	The deviatoric component of $\bten{\Sigma}$ provides the shear viscosity,  $ \mu = \mu_f + \varphi \mu_p $, where $ \varphi \mu_{p} = \Sigma_{p\,xy}/\dot{\gamma} $ is due to the suspended microswimmers and 
		$ \varphi = n a^3 $
	is proportional to their volume fraction with $n$ being the particle density.
	%
	%
	Note that the shear viscosity $\mu_f$ of the second-order fluid does not depend on the shear rate, while $\mu_p$ is dependent on it.
	To evaluate $\mu_p$, we need to average over the stresslets $\bten{\Pi} (\IB{p})$  generated by the suspended particles. Using the orientational distribution, we obtain
	\begin{equation}\label{ensemble}
		\bten{\Sigma}_{p} = n \ang{ \bten{\Pi}} = n\int_{S} \bten{\Pi} (\IB{p}) \psi(\IB{p}) \text{d}\IB{p}, 
	\end{equation}
	where $\bten{\Pi} (\IB{p})$ is the sum of three contributions \cite{saintillan2010dilute}, which give the following stress tensors:
	$ \bten{\Sigma}^{A}_{p} = n   \sigma^{*}  \left[\ang{\IB{pp}} - \bten{I} /3\right] /8\pi $ due to activity,
	$ \bten{\Sigma}^{T}_{p} = 3 n k_B T   \left[\ang{\IB{pp}} - \bten{I} /3\right] $ due to thermal reorientations \cite{batchelor1970}, and
	$ \bten{\Sigma}^{F}_{p} $ due to the resistance of passive particles to shear. 
	For Newtonian fluids, the latter was first derived
	by Einstein \cite{einstein1906new} for spherical particles, and then later generalized to elongated particles
	by Hinch and Leal \cite{hinch1975constitutive,hinch1976constitutive}.
	For the second-order fluid, 
	we follow Férec \emph{et al.}\ \cite{ferec2017steady} and approximate it using the Geisekus form,
	$\bten{\Sigma}^{F}_{p} = -\mu_f n l^{3} A \dot{\gamma}  \overset{\nabla}{\ang{\IB{pp}}}/2 $, where $ A = \pi/6 \ln(2\lambda)$ is a shape factor 
	to access the friction of a long slender body and $ \overset{\nabla}{\ang{\IB{pp}}} $ is the  upper-convected time derivative of the second moment of $\psi(\IB{p})$. 
	The expression for $ \bten{\Pi}^{F} $ was originally derived for dumbbells suspended in a Newtonian fluid \cite{birdTWO1987dynamics}. 
	As in Ref.\ \cite{ferec2017steady}, we use it here as an approximation for the stress response of particles in a weakly viscoelastic fluid, as there are no closed form expressions available. 
	The consequences
	of the second-order fluid come in through the orientational dynamics $ \dot{\IB{p}}$ 
	in Eq.\ (\ref{EOM}) and further terms in the upper-convected derivative (see `Methods: Rheology'). 
	%
	The expressions for the three contributions to the particle stress tensor
	are pertinent to studies on slender rods and fibres (\emph{i.e.,} $ \Lambda \to 1 $),  and henceforth, we will be focusing on spheroids with larger aspect ratios
	($ \lambda=10 $).
	Using Eq.\ (\ref{ensemble}) in the definition of the particle-induced contribution to shear viscosity, $ \varphi \mu_{p} = \Sigma_{p\,xy}/\dot{\gamma} $, 
	together with our 
	characteristic parameters, 
	we obtain
			\begin{equation} \label{visc_res}
				\frac{\mu_{p}}{\mu_f} = 
				-4 A \overset{\nabla}{  \ang{\IB{pp}}  }_{xy}+ \frac{8}{ \text{Pe}_{f} } \left( 6 \tau D_r A - \frac{\text{Pe}_a}{8\pi} \right) \left[ \ang{\IB{pp}} - \frac{\bten{I}}{3} \right]_{xy}. 
			\end{equation}
		%
		%
		\noindent
		We use numerical solutions of Eq.\ (\ref{Smol}) for the orientational distribution $\psi(\IB{p})$ to evaluate $\mu_p$ for
		varying $\mathrm{Pe}_f$ and $\mathrm{Pe}_a$. We also match the numerical values of $\mu_p$ with an expression, which
		we obtain using the analytic form of $\psi(\IB{p})$ from Eq.\ (\ref{pert_Smol}) in the limit of small $ \text{Pe}_{f} $ (the derivation  is provided in `Methods: Rheology'). In Supplementary Figure 4, we also show that our results agree 
		with Saintillan \cite{saintillan2010dilute} in the Newtonian limit.
		
		%

		

		Figure\ \ref{fig: viscosity}a shows the 
		particle-induced contribution to shear viscosity normalized by the bare fluid viscosity $\mu_f$,
		which is plotted versus shear strength $\text{Pe}_f$. We begin with discussing suspensions in a Newtonian fluid (dashed lines) \cite{saintillan2010dilute}.
		When a 
		suspension of passive rods (black) is sheared weakly ($ \text{Pe}_{f} \ll 1 $), the orientational microstructure is governed by Eq.\ (\ref{pert_Smol}) with $\text{De} = 0$ and resembles  Fig.\ \ref{fig: microstructure}a. 
		Here, the 
		rods are aligned along the extensional axis of the applied shear flow and resist shearing ($\mu_p > 0$), which enhances viscosity.
		%
		For a suspension of pushers (brown dashed) in the weak-shear regime, the microstructure still resembles Fig.\ \ref{fig: microstructure}a,
		but now the extensile force dipoles support the elongational part of the shear flow, which results in $\mu_p < 0$ so that 
		the total shear viscosity is smaller than $\mu_f$.
		Conversely, 
		pullers (green dashed) with their contractile force dipoles additionally resist shear flow and thereby enhance viscosity.
		Note that, since the microstructure of active suspensions in a Newtonian fluid
		is unchanged by the hydrodynamic signature of microswimmers, the magnitude of the activity contribution
		to $\mu_p$
		is identical for pushers and pullers.
		As the shear rate increases, the magnitude of 
		$\mu_p$ for both passive and active rods reduces, which resembles a typical shear-thinning/thickening behaviour.
		For passive particles the microstructure is similar to Fig.\ \ref{fig: microstructure}e.
		The alignment close to the flow axis explains why the resistance of particles to the applied shear flow
		is reduced.
		%
		For active particles, the activity-induced flow  becomes  less and less important with increasing shear rate as quantified by the inverse proportionality to $ \text{Pe}_{f} $ in Eq. (\ref{visc_res}).

		\begin{figure}[t]
			\centering
			\includegraphics[width=1.0\textwidth]{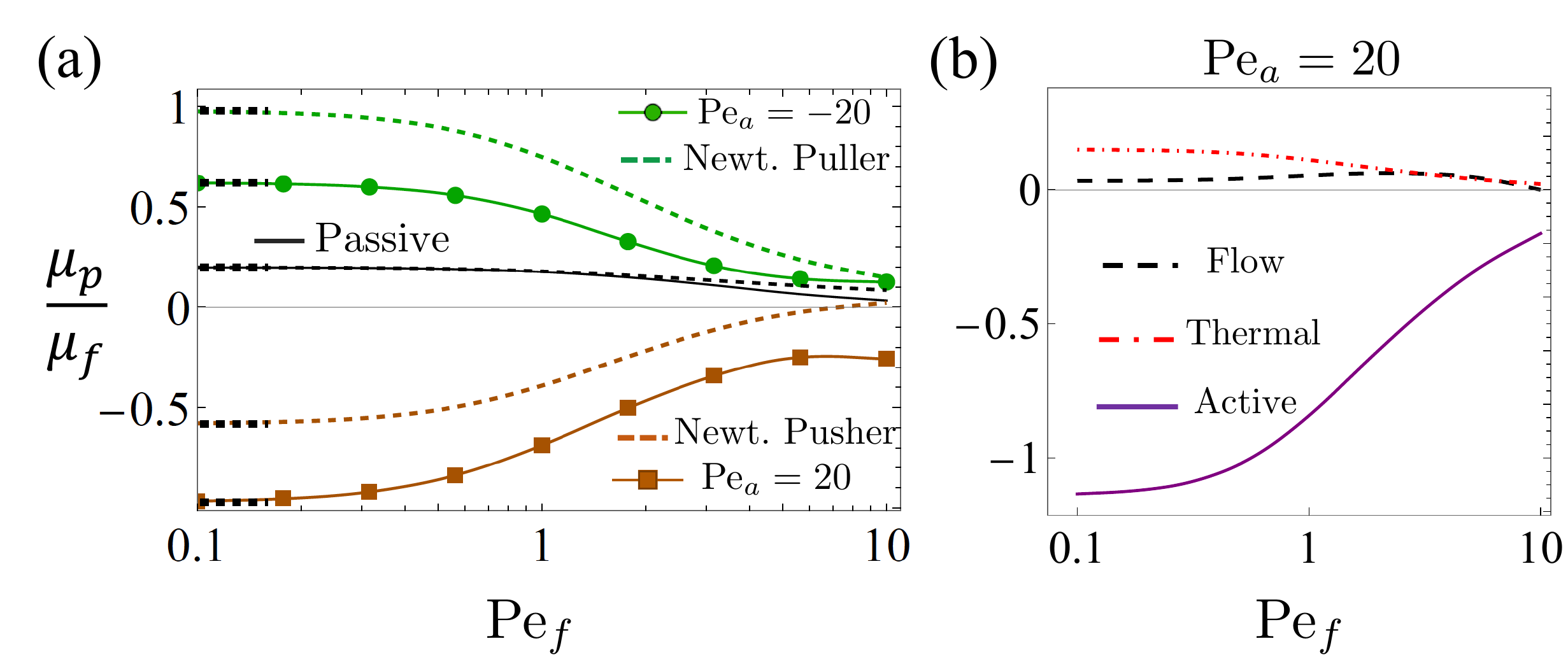}
			\includegraphics[width=1.0\textwidth]{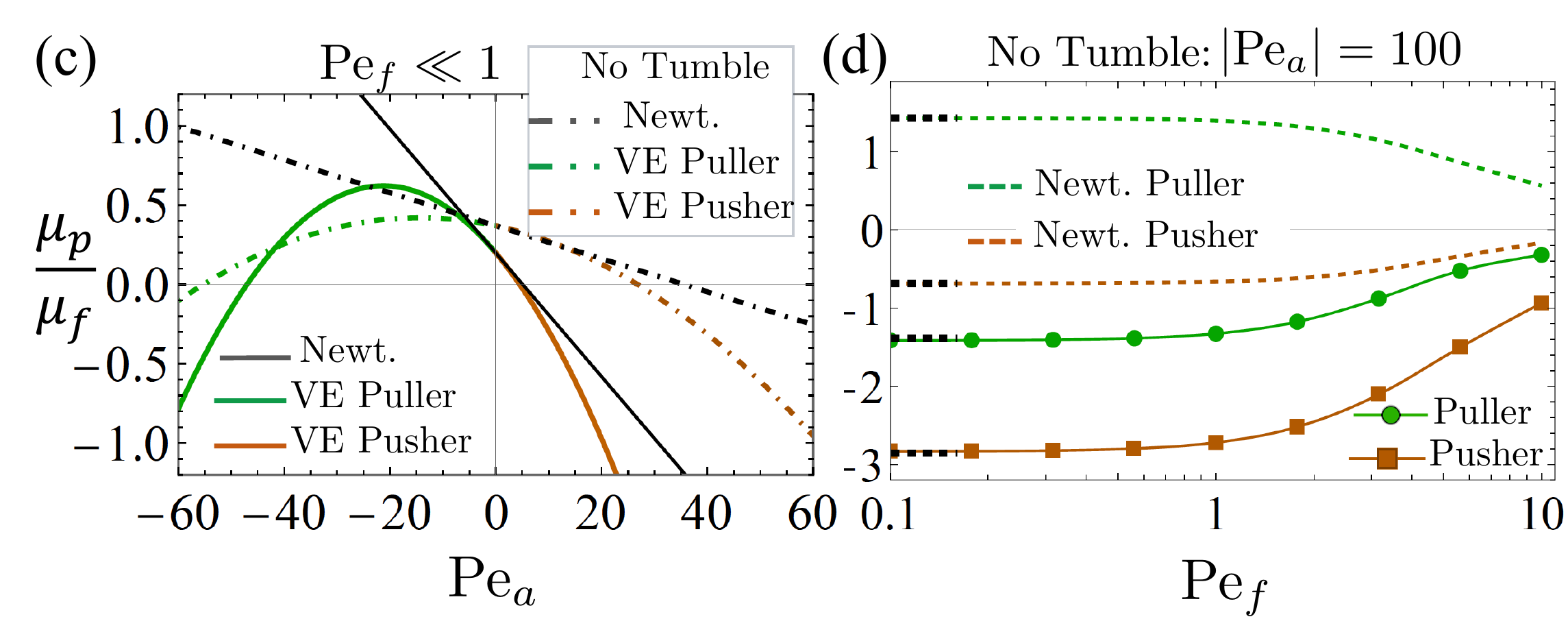}
			\caption{
				\textbf{Particle-induced contribution to viscosity 
				} $ \mu_{p}/\mu_{f} $ in a weakly viscoelastic (VE) fluid:
				(a) plotted \emph{vs.} shear rate (characterized by the flow Péclet number $ \text{Pe}_f$) 	for pusher ($ \text{Pe}_{a}>0 $), puller ($ \text{Pe}_{a}<0 $), and a passive rod ($ \text{Pe}_a = 0 $) with solid lines.
				Dashed lines correspond to the Newtonian case.
				(b) Contributions to particle-induced viscosity for activity Péclet number $ \text{Pe}_{a}=20 $.
				(c) $ \mu_{p}/\mu_{f} $ \emph{vs.} activity $\text{Pe}_{a}$ in the  $ \text{Pe}_{f} \ll 1 $ limit for tumbling and non-tumbling active rods.
				(d) $ \mu_{p}/\mu_{f} $ \emph{vs.} shear rate $ \text{Pe}_f$
				for non-tumbling active rods with large persistence length, $ |\text{Pe}_{a} | = 100 $.
				Dotted black lines near $ \text{Pe}_f = 0.1 $ in (a,d) correspond to the analytical results for $ \text{Pe}_{f} \ll1 $. 
				Other parameters: Deborah number $ \text{De}=0.1 $, the active and passive viscoelastic coefficients of Eq. (\ref{EOM}) are evaluated to be  $ \alpha_{1}=5.92, \; \alpha_{2}=-0.91, \; \beta_{1}=0.63, \; \beta_{2}=1.18 $ for aspect ratio $ \lambda=10 $.	
				%
		}
		\label{fig: viscosity}
	\end{figure}

	Now, we discuss the results for the second-order fluid (SOF) in Fig. \ref{fig: viscosity}a (solid lines). For the suspension of 
	passive rods (black), we do not find an appreciable deviation from the Newtonian case until $ \text{Pe}_{f} \approx 10 $, similar to Refs. \cite{cohen1987orientation,chung1987orientation}.
	However, for active suspensions  the modifications are substantial and qualitatively different for pushers (brown) and pullers (green).
	For pushers, the reduction in viscosity is further enhanced compared to Newtonian fluids, whereas the viscosity enhancement of pullers is  reduced. 
	{For a fixed viscosity of $ \mu_{f} $ and a volume fraction of $ \varphi = 0.1 $, a pusher suspension of $ \text{Pe}_{a} =20 $ predicts a $ 10\% $ viscosity reduction, whereas in a Newtonian fluid the reduction is $ 6 \% $.}
	We understand this {viscosity reduction by} using the orientational distributions in Fig. \ref{fig: microstructure}:
	pushers are even more aligned in the extensional quadrant of the applied shear flow 
	(Fig. \ref{fig: microstructure}b,d,f), which explains their further increased support of shear flow. 
	In contrast, pullers are less aligned compared to the Newtonian case (Fig. \ref{fig: microstructure}c,d,g) and, therefore, oppose the shear flow to a reduced extent.
	Figure \ref{fig: viscosity}b for pushers clearly shows that the active stress ($ \bten{\Sigma}^{A} $) dominates and hence is primarily responsible for the observed viscous response.

	%
	In the limit of shear rate much smaller than tumbling rate, $\text{Pe}_{f} \ll 1$
	, we can calculate $\mu_p/\mu_f$ analytically using Eq.\ (\ref{pert_Smol}) (see `Methods: Rheology')
	and obtain:
	\begin{align}
		\label{eq.mup}
		\frac{\mu_{p}}{  \mu_{f} } =  	&\frac{ 4A(5-3\Lambda) }{15} +   \frac{  48 A \tau D_{r} - \text{Pe}_{a} / \pi}{5\left( 1+ 6 \tau D_r \right)}  \left( \Lambda + \frac{\text{De} \, \text{Pe}_{a} \alpha_{1}}{8\pi}  \right)  \nonumber \\
		& -  \frac{\text{De} \, \text{Pe}_{a} A \alpha_{1}}{10\pi}. 
	\end{align}
	We elaborate the result here since the influence of passive and active rods on the shear viscosity is the strongest in this limit.
	%
	In Fig.\ \ref{fig: viscosity}c we plot $\mu_p/\mu_f$ \emph{vs.} $\text{Pe}_{a}$ for tumbling microswimmers (solid lines).
	As Eq.\ (\ref{eq.mup}) shows, elasticity in the fluid adds a quadratic dependence in $\text{Pe}_{a}$, while in a Newtonian fluid the dependence is only linear \cite{saintillan2010dilute}.
	%
	%
	For 
	all $\text{Pe}_{a}$,
	except a small region close to
	$\text{Pe}_{a} =0$, the values of $\mu_p/\mu_f$
	remain below 
	those obtained in the Newtonian limit. 
	Thus,
	fluid elasticity reduces the shear viscosity for both pusher and puller suspensions.
	Note that, as we increase the activity of pullers ($ \text{Pe}_{a} \lesssim -50 $), 
	they too can reduce the shear viscosity ($\mu_p/\mu_f < 0$) like pushers.
	%
	This is a consequence of the orientational
	distribution  shown in Fig. \ref{fig: microstructure}j; pullers with large activity align preferentially in the compressional 
	quadrant of the shear flow and thereby also support the shearing fluid similar to pushers aligned in the extensional quadrant.

	As described earlier, we can treat the case of non-tumbling microswimmers by replacing $ \tau $ by $ D_{r}^{-1} $ in the definitions of $ \text{Pe}_{f} $, $ \text{Pe}_{a} $, and $ \text{De}$. 
	With these parameters, Eq.\ (\ref{eq.mup}) can be formulated in the limit $\tau \rightarrow \infty$ to show that the viscosity contribution of non-tumbling active rods
	also follows a quadratic dependence in $ \text{Pe}_{a} $ (Fig.\ \ref{fig: viscosity}c, dashed lines). 
	Non-tumbling microswimmers can exhibit high persistence lengths.
	Thus, in Fig.\ \ref{fig: viscosity}d we show for  $|\text{Pe}_{a}| =100 $
	that their contribution to viscosity is negative over a wide range of shear rates  for both pushers and pullers in a second-order fluid.
	Hence, we find that the elasticity of a fluid always results in a reduced total viscosity as compared to suspensions in a Newtonian fluid {of identical base viscosity ($ \mu_f $)}.

	\section*{Discussion}
	In this article we show how activity influences the dynamics of a sheared microswimmer suspension in a viscoelastic fluid at an individual level and in the bulk.
	At the individual level, the orientational dynamics of passive rods is well known from Jeffery's orbits in Newtonian shear 
	\cite{jeffery1922motion} and `log-rolling' orbits in viscoelastic shear flow \cite{leal1975slow,brunn1977slow}.
	Our analytical result  [Eq.\ (\ref{EOM})], 
	derived for elongated active spheroids in a second-order fluid, demonstrates how the active flow field of a swimmer modifies the orientational dynamics. 
	Extensile swimmers like \emph{E.coli} (pushers) drift to the shear plane and align at an angle to the flow direction. For contractile swimmers like \emph{C. reinhardtii} (pullers), activity effectively lowers 
	their aspect ratio.
	With increasing dipole strength, pullers show log rolling,
transition to shear-plane rotation, and ultimately align at an angle against the flow direction. 
The latter occurs for strong pullers at dipole strengths  relevant to artificial swimmers with large propulsion speed  \cite{ren20193d,aghakhani2020acoustically}.

To demonstrate the impact of the individual swimmer dynamics on the bulk rheological behaviour, we employ the Smoluchowski equation to evaluate the orientational probability distribution of  an ensemble of  active spheroids
called the suspension microstructure \cite{hinch1975constitutive}. 
Accounting for tumbling and rotary diffusion, we find that the microstructure {is sensitive to} the hydrodynamic signature of the swimmer unlike suspensions in Newtonian fluids \cite{hatwalne2004rheology,saintillan2010dilute,nambiar2017stress}.
Pushers align more strongly in the extensional quadrant of the applied shear flow, while the alignment of pullers is significantly weaker.
%
This activity-specific microstructure 
significantly modifies the effective shear viscosity compared to Newtonian 
fluids \cite{lopez2015turning}.
In particular, the viscosity reduction of a dilute pusher suspension is more pronounced under viscoelastic shear flow,
while the viscosity enhancement of pullers is weaker. Thus, the activity of a microswimmer contributes in 
two ways to the rheology of swimmer suspensions in a viscoelastic fluid; directly through its active stresses and indirectly by coupling to
the elasticity of the fluid and thereby influencing the orientational microstructure. 
In particular, in the weak-shear limit the particle-induced contribution to viscosity scales quadratically with the 
activity $ \text{Pe}_{a} $.
%
%
%
In total, we find that the elasticity of a second-order fluid always reduces the total viscosity of the microswimmer suspension{, as compared to a Newtonian fluid of identical base viscosity}.
Especially for pushers, this {might help} to reach the regime of superfluidity at lower volume fractions compared to Newtonian fluids. 
{We note that superfluidity is reported to be also associated with onset of collective motion \cite{guo2018symmetric,aranson2022bacterial}, and requires further analysis at higher microswimmer densities.}

Most biological fluids are viscoelastic. We presented a first systematic study of the individual dynamics and the bulk rheology of microswimmers suspended in such fluids. Earlier investigations on active suspensions in quiescent \cite{bozorgi2011effect} and vortical viscoelastic flows \cite{ardekani2012emergence} assumed conventional Newtonian Jeffery orbits.  In the light of our results, 
these studies on the collective behaviour might need to be revisited. 
Our investigations makes several
assumptions to address the complexity of microbial flows,  which offers ample opportunities for future developments. Biological fluids can be more complex and exhibit shear-thinning/-thickening properties or several characteristic relaxation times, which requires more evolved modeling using, for example, the 
FENE-P or Giesekus model.
%
%
Furthermore, mucus besides being significantly shear-thinning has relaxation times larger than 1 second, which is of the order of
the bacterial mean free tumble time \cite{li2021microswimming,spagnolie2022swimming,mucus_lai2009micro}.
Thus, noise becomes non-Markovian and memory needs to be incorporated in the stochastic description, for example, within
a generalized Langevin equation \cite{narinder2018memory}.



\setcounter{subsection}{0}

\section*{Methods}
\label{sec.methods}

\subsection{Hydrodynamic Model}
The hydrodynamics around the spheroidal particle is governed by the mass and momentum conservation as
$ \nabla \cdot \IB{V} = 0, \;  \nabla \cdot \bten{T}=0. $
Here $ \bten{T} $ is the total stress tensor defined as $ \bten{T} =-P \, \bten{I} + 2 \, \bten{E} + \text{Wi} \, \bten{S} $ \cite{bird1987dynamics}, where 
$
\bten{S} =	  4 \bten{E}\cdot \bten{E} + 2 \delta \overset{\Delta}{\bten{E}}
$. The lower-convected derivative is defined as
\begin{equation}
\overset{\Delta}{\bten{E}} = \left[\frac{\partial\bten{E}}{\partial t}  + \left( \IB{V} \cdot \IB{\nabla} \right) \bten{E} + \bten{E} \cdot \IB{\nabla} \IB{V} + \IB{\nabla V}^{T} \cdot \bten{E}  \right] \,.
\end{equation}
Since we consider a steady shear flow and a steadily active swimmer, we can disregard the temporal derivative in the above stress tensor.
In particle frame of reference, the boundary condition on its surface is the no-slip condition
$ 	\IB{V} = \IB{\Omega}^{p} \times \IB{r} $,
where the rotational velocity $ \IB{\Omega}^{p} $ is currently unknown and will be evaluated by using the torque-free condition.
We split the velocity field into the disturbance field $\IB{v}$ and background flow field $\IB{V}^{\infty}  = \bten{E}^{\infty} \cdot \IB{r} + \IB{\Omega}^\infty \times \IB{r} $ \emph{i.e.,} $ \IB{V}=\IB{v}+\IB{V^{\infty}} $.
The equations governing the disturbance flow field are 
\begin{eqnarray}  \label{GE_supp}
\nabla \cdot \IB{v} &=& 0, \quad  - \nabla p + \nabla^{2} \IB{v} = - \text{Wi} \, (\nabla \cdot \bten{s}), \text{ where} \nonumber \\ 
\bten{s} &=& 4 (\bten{e} \cdot \bten{e} + \bten{w}) + 2 \, \delta(\overset{\Delta}{\bten{e}}+\overset{\Delta}{\bten{w}}).
\end{eqnarray}
Here $ \bten{e} $ is the rate of strain tensor for the disturbance flow $ (\nabla \IB{v} + \nabla \IB{v}^{\dagger})/2 $, whereas $ \bten{w} $, $ \overset{\Delta}{\bten{e}} $ and $ \overset{\Delta}{\bten{w}} $ are the different constituents of the polymeric tensor ($ \bten{s}$) due to the disturbance flow field.
Specifically, $ \bten{w} $ is the interaction tensor defined as $  \bten{E}^{\infty}\cdot \bten{e} + \bten{e} \cdot  \bten{E}^{\infty} $, arising from the interaction between background flow and disturbance field;
and $ \overset{\Delta}{\bten{w}} $ is the lower-convected derivative of $ \bten{e} $ and $  \bten{E}^{\infty} $ with respect to $ \IB{V}^{\infty} $ and $ \IB{v} $, respectively:
\begin{eqnarray}
&\overset{\Delta}{\bten{w}}  = {\IB{V}^\infty} \cdot \nabla \bten{e} + \bten{e} \cdot \nabla {\IB{V}^\infty}^{\dagger} + \nabla {\IB{V}^\infty} \cdot \bten{e} \\
&\qquad \qquad \qquad+  \IB{v} \cdot \nabla  \bten{E}^{\infty}+  \bten{E}^{\infty}\cdot \nabla \IB{v}^{\dagger} + \nabla \IB{v} \cdot  \bten{E}^{\infty}. \nonumber
\end{eqnarray}
The boundary conditions of the disturbance field at the particle surface and far away are
\begin{equation}
\IB{v} =  \Delta \IB{\Omega} \times \IB{r}  -  \bten{E}^{\infty} \cdot \IB{r}  \quad \mbox{at\ } \IB{r} \in S \quad \text{and}  \quad   \IB{v} \to \IB{0} \quad \mbox{as\ } \; \IB{r} \to \infty,
\end{equation}
respectively. Here $  \Delta \IB{\Omega}  $ is the difference in particle angular velocity and background vorticity ($   \IB{\Omega}^{p} - \IB{\Omega}^{\infty} $). 


For small values of $ \text{Wi} $, the disturbance field variables are expanded as $ 	f=f^{(0)} + \text{Wi} \; f^{(1)} + \cdots. $
%
Here, $ f $ is a generic field variable that represents velocity ($ \IB{v} $), pressure ($ p $), and angular velocity ($ \IB{\Omega}^{p} $). 
We substitute this expansion in the Eq. (\ref{GE_supp}) and obtain the $ O(1) $ Stokes problem as
\begin{eqnarray}\label{GE}
\nabla \cdot \IB{v}^{(0)} &=& 0, \quad  - \nabla p^{(0)} + \nabla^{2} \IB{v}^{(0)} = 0
\end{eqnarray}
with the boundary condition at particle surface:
\begin{equation}
\IB{v}^{(0)} =  \Delta \IB{\Omega}^{(0)} \times \IB{r}  -  \bten{E}^{\infty} \cdot \IB{r}  \quad \mbox{at\ } \IB{r} \in S,
\end{equation}
and a decaying condition at infinity ($ \IB{v}^{(0)} \to 0 $).
Using the finite multipole expansion around the spheroid \cite{chwang1975hydromechanics,einarsson2015rotation,abtahi2019jeffery}, we obtain the disturbance velocity for a passive spheroid at $ O(1) $ as 
{
\small
\begin{align}\label{dist}
	v^{(0)}_{i}  &=  \mathcal{Q}_{ij,k}^{R} \varepsilon_{jkl} \Bigl\{  \left[  \mathbb{A}^{R} p_{l} p_{m} + \mathbb{B}^{R} (\delta_{lm}-p_{l} p_{m})  \right] \Delta\Omega_{m}  
	\nonumber \\
	&  \!\!\!\!\!\!	+\mathbb{C}^{R} \varepsilon_{lmn} p_{m} \mathsf{E}^{\infty}_{no} p_{o} \Bigr\} 
	\nonumber \\
	& \!\!\!\!\!\! + \left(\mathcal{Q}_{ij,k}^{S} + \chi \mathcal{Q}_{ij,llk}^{Q}\right) \Bigl\{  \left[ \mathbb{A}^{S} {\mathsf{p}^{A}_{jklm}} + \mathbb{B}^{S} {\mathsf{p}^{B}_{jklm}} + \mathbb{C}^{S} {\mathsf{p}^{C}_{jklm}}  \right]  \mathsf{E}_{lm}^{\infty}
	\nonumber \\
	&\!\!\!\!\!\!  - \mathbb{C}^{R} \left( \varepsilon_{jlm} p_{k}p_{m} +\varepsilon_{klm} p_{j} p_{m} \right) \Delta \Omega_{l} \Bigr\} . 	
\end{align}
}
\noindent
Here $ \mathcal{Q}$  represent the spheroidal multipoles \emph{i.e.} integral representation of fundamental singularities spread on a line extending from one foci to another ($ -c $ to $ c $, where $ c=\sqrt{\lambda^{2}-1}/\lambda $);
$ \mathcal{Q}^{R}, \; \mathcal{Q}^{S} $ represent rotlet and stresslet around the spheroid, respectively.
Here $ \chi = \frac{1}{8(\lambda^{2}-1)} $ represents the strength of higher order quadrupolar field ($ \mathcal{Q}_{ij,llk}^{Q} $). Since we focus on elongated particles, we exclude this component for computational ease ($ \lambda=\lbrace 5,10 \rbrace $ correspond to $ \chi = \lbrace0.005,0.001 \rbrace $).
Following Einarsson \emph{et al.} \cite{einarsson2015rotation}, we simplify these multipoles as
\begin{subequations}\label{spheroid_multipole}
\begin{align}
	\mathcal{Q}_{ij,k}^{R} &= \left( \delta_{ik} r_{j} - \delta_{ij} r_{k} \right) J_{3}^{0} + \left( \delta_{ij} p_{k} - \delta_{ik} p_{j} \right)J_{3}^{1} 
	\\
	\mathcal{Q}_{ij,k}^{S} &= \delta_{jk} r_{i} J_{3}^{0} - \delta_{jk} p_{i} J_{3}^{1} - 3  r_{i} r_{j} r_{k} J_{5}^{0}  \\ \nonumber
	&+ 3(p_{i}r_{j}r_{k} + p_{j}r_{i} r_{k} +p_{k} r_{i} r_{j} ) J_{5}^{1} \\ \nonumber
	&- 3(r_{i}p_{j}p_{k} + r_{j}p_{i} p_{k} +r_{k} p_{i} p_{j} ) J_{5}^{2} 
	+ 3 p_{i} p_{j} p_{k} J_{5}^{3}.
\end{align}
\end{subequations}
Here $ I $ and $ J $ represent various integrals defined as
\begin{align}
I_{a}^{b} = \int_{-c}^{c}  \frac{\xi^{b}}{|\IB{r}-\xi \IB{p}|^{a}} \text{d}\xi ; \; J_{a}^{b} = c^{2} I_{a}^{b}  -  I_{a}^{b+2}. 
\end{align}
In Eq. (\ref{dist}), the fourth order orientation tensors ($ \mathsf{p}^{A} $, $ \mathsf{p}^{B} $, $ \mathsf{p}^{C} $) and other coefficients ($ \mathbb{A}, \mathbb{B}, \mathbb{C} $) are described in the Supplementary Note 1.

At $ O(1) $, we add the activity using the far-field descriptions, which consists of a force-dipole (FD), source-dipole (SD), rotlet-dipole (RD), and force-quadrupole (FQ) \cite{spagnolie_lauga_2012}:
{
\small
\begin{subequations}\label{active_fields}
	\begin{equation}
		\IB{v}^{(0)}_{FD}	= \sigma_{FD} \left(\frac{-  \IB{r} }{r^{3}} +  \frac{3  \IB{r} \left(\IB{r} \cdot \IB{p}  \right)^{2} }{r^{5 }} \right)
	\end{equation}
	\begin{equation}
		\IB{v}^{(0)}_{SD}	= \sigma_{SD} \left( \frac{ 3 \IB{rr}}{r^{2}}  -\bten{I} \right)  \frac{\IB{p}}{2 r^{3}}
	\end{equation}
	\begin{equation}
		\IB{v}^{(0)}_{RD}	= \sigma_{RD} \left[  \frac{ 3 \IB{p}\cdot \IB{r} }{r^{5}}   \left( \IB{p} \times \IB{r} \right)  \right]
	\end{equation}
	\begin{equation}
		\IB{v}^{(0)}_{FQ}	= \sigma_{FQ}  \left\lbrace \frac{\IB{p}}{r^{3}} -  \frac{3}{r^{5}} \left[ 3(\IB{p}\cdot \IB{r} )  \IB{r}  + (\IB{p}\cdot \IB{r})^{2} \IB{p}  - \frac{5 (\IB{p}\cdot \IB{r})^{3} \IB{r} }{r^{2}} \right]   \right\rbrace
	\end{equation}
\end{subequations}
}



At $ O(\text{Wi}) $, the governing equation is
\begin{eqnarray}\label{dist_Wi}
\nabla \cdot \IB{v}^{(1)} &=& 0, \quad  - \nabla p^{(1)} + \nabla^{2} \IB{v}^{(1)} = -  \, (\nabla \cdot \bten{s}^{(0)}), \text{ where} \nonumber \\ 
\bten{s}^{(0)} &=& 4 (\bten{e}^{(0)} \cdot \bten{e}^{(0)} + \bten{w}^{(0)}) + 2 \, \delta(\overset{\Delta}{\bten{e}}^{(0)}+\overset{\Delta}{\bten{w}}^{(0)}),
\end{eqnarray}
The boundary condition at the particle surface being
\begin{equation}\label{BC_Wi}
\IB{v}^{(1)} = \IB{U}_{p}^{(1)} + \IB{\Omega}_{p}^{(1)} \times \IB{r}  \quad \mbox{at\ } \IB{r} \in S,
\end{equation}
with a decaying condition at infinity ($ \IB{v}^{(1)} \to 0 $).
Conventionally, a solution to Eqs. (\ref{dist_Wi}) and (\ref{BC_Wi})  (\emph{i.e.,} $ \IB{v}^{(1)} $) is sought, which, on the implementation of torque-free condition, reveals the modification of Jeffery orbits ($ \IB{\Omega}_{p}^{(1)} $).
We employ the reciprocal theorem \cite{elfring2015theory,abtahi2019jeffery} to avoid solving for the first order flow field and directly obtain the rotation velocity as
\begin{equation}\label{reciprocal}
\IB{\Omega}_{p}^{(1)} \cdot \IB{T}^{t} = \int_{V_{f}} \bten{s}^{(0)} : \nabla \IB{v}^{t} \; dV. \quad  
\end{equation}
Here, superscript $ t $ refers to the test problem where the particle rotates at a unit velocity $ \IB{\Omega}^{t} = \IB{e}_{i} $, where $ i $ corresponds to either of the three Cartesian coordinates.
The test torque is:
\begin{equation}
\IB{T}^{t} = \bten{R} \cdot \IB{\Omega}^{t}, \, \mbox{where\ } \bten{R} =  \frac{64 \pi c^{3}}{3}  \left[  \mathbb{A}^{R} \IB{pp} + \mathbb{B}^{R}(\bten{I}-\IB{pp})  \right] 
\end{equation}
We solve Eq.\ (\ref{reciprocal}) for three test field torques (along the three Cartesian coordinates) and obtain the following system of equations:
$ \bten{R}^{\dagger} \cdot \IB{\Omega}_{p}^{(1)} = \left\{ \mathcal{I}_{1}, \mathcal{I}_{2}, \mathcal{I}_{3} \right\} $,
where $ \dagger $ represents transpose and $ \mathcal{I}_{i} $ is the solution of volume integral (\ref{reciprocal}) for the test field in $ i^{\text{th}} $ unit vector. We further simplify the above relation by noting that $ \bten{R} $ is a symmetric matrix and
$ \dot{\IB{p}}^{(1)}=\IB{\Omega}_{p}^{(1)} \times \IB{p} $:
\begin{equation}\label{pdot}
\dot{\IB{p}}^{(1)}	= \left( \bten{R}^{-1} \cdot \left\{ \mathcal{I}_{1}, \mathcal{I}_{2}, \mathcal{I}_{3} \right\} \right)	\times \IB{p}
\end{equation}
Eq.\ (\ref{pdot}) can be evaluated over a discretized angular grid, where each point requires solving the volume integral in Eq. (\ref{reciprocal}). 
The polymeric stress  therein ($ \bten{s}^{(0)} $) is given by Eq. (\ref{GE_supp}) and is entirely dependent on $ O(1) $ disturbance flow fields \emph{i.e.,} passive disturbance in Eq.\ (\ref{dist}) and active disturbance in Eq. (\ref{active_fields}).
We find that out of all active fields, only force-dipole ($ \sim r^{-2} $) provides a modification at the current order of approximation; the rest decays in odd powers of distance ($ \sim r^{-3} $), and due to antisymmetry, give zero contribution to the volume integral at $ O(Wi) $.

%
\subsection{Orientation dynamics}
We evaluate 
$ \dot{\IB{p}}^{(1)} $ analytically by noting that it stems from the leading order polymeric stress [specifically $ \bten{s}^{(0)} $ in Eq.\ (\ref{reciprocal})];
its form in Eq.\ (\ref{GE_supp}) suggests that the modification for a passive particle will be quadratic in the flow gradient tensor, which can be decomposed in symmetric $ \bten{E}^{\infty} $ and antisymmetric $ \bten{O}^{\infty} $ (rotation-rate tensor) components.
Following Einarsson \emph{et al.}\ \cite{einarsson2015rotation}, this can be written in the general form as:
\begin{equation}\label{p_passive}
\dot{p}_{i}^{(1,P)} = \mathsf{K}_{ijklm}^{(1)} \mathsf{E}^{\infty}_{jk} \mathsf{E}^{\infty}_{lm}  +  \mathsf{K}_{ijklm}^{(2)} \mathsf{E}^{\infty}_{jk} \mathsf{O}^{\infty}_{lm}  + \mathsf{K}_{ijklm}^{(3)} \mathsf{O}^{\infty}_{jk} \mathsf{O}^{\infty}_{lm},
\end{equation}
where superscript $ P $ denotes passive contribution.
The coefficients of the fifth-order tensor $ \bten{K} $ are composed of all possible permutations of the orientation vector $ \IB{p} $ with $ \bten{E}^{\infty} $ and $ \bten{O}^{\infty} $:
\begin{align}
\mathsf{K}_{ijklm}^{(i)} = \sum_{n=5!} \left( \beta_{n}^{[1]} p_{n_{1}} p_{n_{2}} p_{n_{3}} p_{n_{4}} p_{n_{5}} + \beta_{n}^{[2]} p_{n_{1}} p_{n_{2}} p_{n_{3}} \delta_{n_{4}n_{5}}  \right.
\nonumber	\\ 
\left.
+ \beta_{n}^{[3]} p_{n_{1}} \delta_{n_{2}n_{3}} \delta_{n_{4}n_{5}}  \right) \nonumber.
\end{align}
Here $ \beta $ represents  the unique coefficients for all the $ 5! $ terms.

When activity is added in the hydrodynamics, the form of polymeric stress tensor in Eq. (\ref{GE_supp}) suggests that the gradients of disturbance velocity (directed with $ \IB{p} $) will multiply with gradients of passive disturbance (which originate from $ \bten{E}^{\infty} $ and $ \bten{O}^{\infty} $) to yield further modification to $\dot{\IB{p}}^{(1)}$.
Thus, the contribution resulting from interaction of active disturbances ($ \IB{p} $) with background flow ($ \bten{E}^{\infty} $ and $ \bten{O}^{\infty} $) takes the following form at $ O(\text{Wi}) $:
\begin{equation}\label{p_active}
\dot{p}_{i}^{(1,A)} = \sigma \left[  \mathsf{L}_{ijk}^{(1)} \mathsf{E}^{\infty}_{jk}  +  \mathsf{L}_{ijk}^{(2)} \mathsf{O}^{\infty}_{jk} \right]  ,
\end{equation}
where coefficients of the third-order tensor $ \bten{L} $ are composed of the following combinations
\begin{equation}
\mathsf{L}_{ijk}^{(i)} = \sum_{n=5!} \left( \alpha_{n}^{[1]} p_{n_{1}} p_{n_{2}} p_{n_{3}} + \alpha_{n}^{[2]} p_{n_{1}}  \delta_{n_{2}n_{3}}  \right). \nonumber
\end{equation}
Here $ \alpha $ represent the unique coefficients for all the $ 3! $ terms.
We simplify Eq.\ (\ref{p_passive}) and (\ref{p_active}) by using the symmetry arguments and noting that magnitude of $ \IB{p} $ is always unity. We obtain the following six irreducible terms:
\begin{align}\label{pdot_all}
\dot{\IB{p}}^{(1)} &=  \bten{E}^{\infty}: \IB{pp} \left[ \beta_{1} (\bten{I}-\IB{pp}) \cdot ( \bten{E}^{\infty} \cdot \IB{p}  ) +  \beta_{2} \, \bten{O}^{\infty} \cdot \IB{p}   \right] \nonumber  \\ 
& \; + (\bten{I}-\IB{pp})   \cdot   \left[ \beta_{3} (\bten{E}^{\infty} \cdot  \bten{E}^{\infty}) \cdot \IB{p} 
+ \beta_{4}  (\bten{O}^{\infty} \cdot  \bten{E}^{\infty}) \cdot \IB{p}   \right]  \nonumber  \\ 
& \; + \sigma \alpha_{1}  (\bten{I}-\IB{pp})  \cdot \left( \bten{E}^{\infty} \cdot \IB{p} \right) + \sigma  \alpha_{2}  \,  \bten{O}^{\infty} \cdot \IB{p}
.
\end{align}
We calculate the coefficients ($ \alpha_{i} $ and $ \beta_{i} $) by evaluating $ \dot{\IB{p}}^{(1)} $ numerically for six independent orientations ($ \IB{p} $) and solve the system of equations from Eq.\ (\ref{pdot_all}) to extract the coefficients.
We find that $ \beta_{3} $ and $\beta_{4}$ are nearly zero ($ \lesssim 10^{-5} $) and thus neglect them.
We obtain the analytical approximation as
\begin{align}\label{pdot_supp}
\dot{\IB{p}}^{(1)} &=   \left(  \bten{I} -  \IB{pp}  \right) \cdot  \left(   \bten{E}^{\infty} \cdot \IB{p}   \right) \left[   \sigma  \alpha_{1} + \beta_1 \,  \bten{E}^{\infty}: \IB{pp} \right] 
\nonumber \\
& \; + \bten{O}^{\infty} \cdot \IB{p} \left[   \sigma  \alpha_{2} +  \beta_2 \,  \bten{E}^{\infty}: \IB{pp}   \right] ,
\end{align}
We use the above equation in the main text as Eq. (\ref{EOM}), where we write $ \bten{O}^{\infty}\cdot \IB{p}$ as $ \IB{\Omega}^{\infty} \times \IB{p} $.
In Supplementary Note 2A, we show the weak variation of these coefficients with particle aspect ratio $ \lambda $ and the viscometric coefficient $ \delta $.


\subsection{Dynamical analysis}
\subsubsection{Eigenvalue analysis of alignment dynamics} 
Following Bretherton \cite{bretherton1962}, we explain the alignment of a particle in the shear plane by extracting a fundamental matrix solution of Eq. (\ref{EOM}).
For this, we first consider the non-dimensional Jeffery equation in the expanded form as
\begin{align}
\dot{\IB{p}} & 
= \bten{O}^{\infty}\cdot \IB{p} + \Lambda \left[ \bten{E}^{\infty} \cdot \IB{p} - \IB{p} (\bten{E}^{\infty}:\IB{pp}) \right],
\end{align}
which can be represented in an unconserved form that facilitates a fundamental matrix solution \cite{einarsson_thesis}:
\begin{equation}\label{EOM_unconserved}
\dot{\IB{q}} = \left( \Lambda \bten{E}^{\infty} + \bten{O}^{\infty}  \right) \cdot \IB{q}.
\end{equation}
Here $ \frac{\IB{q}(t)}{|\IB{q}(t)|} = \IB{p}(t)  $ and $ |\IB{q}(t)| = \exp[ \Lambda \left( \bten{E}^{\infty} : \IB{pp} \right) t ] $ represents the exponential elongation of $ \IB{q} $. 
We study the angular dynamics of $ \IB{q} $, as its normalization yields back the orientation vector $ \IB{p} $.
The solution to Eq. (\ref{EOM_unconserved}) is
\begin{equation}
\IB{q}(t) = \exp\big[ ( \Lambda \bten{E}^{\infty}+\bten{O}^{\infty}  )t ] \IB{q}(0).
\end{equation}
The eigensystem of the above exponential matrix determines the orbital dynamics of spheroid.
For the two-dimensional shear flow, the eigenvalues of the matrix in the exponent ($ \Lambda \bten{E}^{\infty} + \bten{O}^{\infty} $) are: $ 0, \, \pm  \sqrt{\Lambda ^2-1}/2  $. The non-zero eigenvalues are purely imaginary as $ 0 < \Lambda < 1 $, where $\Lambda=1$ for an infinitely slender particle. 
As a consequence of this imaginary pair, we observe degenerate infinite solutions of orbits in Newtonian fluids (\emph{i.e.,} Jeffery orbits).
For the case of pure elongational flow ($ \bten{O}^{\infty} \equiv 0 $),  the eigenvalues are real: $ 0, \, \pm \Lambda/2 $, which reveals the absence of periodic orbits.
The normalized eigenvector corresponding to the positive eigenvalue determines the equilibrium orientation: $ \left\lbrace {1}/{\sqrt{2}}, \, {1}/{\sqrt{2}}, \, 0  \right\rbrace $ \emph{i.e.,} $ 45 $\textdegree$\,$ or $ 225 $\textdegree$\,$ in the two extension quadrants.

We now use Eq.\ (\ref{EOM}) for the second-order fluid and consider only the active viscoelastic component because the passive effects do not contribute to the alignment [this assumption is later verified by matching the results with numerical integration of complete Eq.\ (\ref{EOM})].
In this case, the equation of motion for the unconserved vector $ \IB{q} $ is
\begin{equation}\label{EOM_avtive_unconserved}
\dot{\IB{q}} = \left[   \bten{E}^{\infty} (\Lambda+ \text{Wi} \sigma \alpha_{1}) + \bten{O}^{\infty} (1+ \text{Wi} \sigma \alpha_{2})  \right] \cdot \IB{q},
\end{equation}
which yields the solution $ q(t) $ as
\begin{align} \label{active_EOM}
	\exp  \big[ \left[  \bten{E}^{\infty}  (\Lambda+ \text{Wi} \sigma \alpha_{1}) + \bten{O}^{\infty} (1+ \text{Wi} \sigma \alpha_{2}) \right] t   ] \IB{q}(0).
\end{align}
We obtain the eigenvalues of the fundamental matrix as
\begin{align}
	0, \, \pm \frac{1}{2} \sqrt{\Lambda ^2 -1 +2 \sigma  \text{Wi} \left(\text{$\alpha_{1} $} \Lambda  -\text{$\alpha_{2} $}\right) + \text{Wi}^{2} \sigma^{2} \left( \alpha_{1}^{2} - \alpha_{2}^{2} \right)} \nonumber .
\end{align}
These eigenvalues turn real when $ \sigma > \frac{ 1 - \Lambda}{\text{Wi} (\alpha_{1} - \alpha_{2})} $ or $ \sigma < \frac{ -1 - \Lambda}{\text{Wi} (\alpha_{1} + \alpha_{2})}  $. 
For instance, for a particle of aspect ratio $ \lambda=5 $, the negative and positive critical values are $ \sigma_{crit} = -0.448/\text{Wi}; \;  0.016/\text{Wi} $, which matches with Fig. \ref{fig:phase} generated from the integration of full Eq. (\ref{EOM}) (see Supplementary Note 2B). 
In the alignment regimes of Fig. \ref{fig:phase}, we further find that the normalized eigenvector corresponding to the positive eigenvalue matches the angle of alignment ($ \phi_{eq} $) obtained via numerical integration.

In addition to validating our results of Fig.\ \ref{fig:phase}, Eq. (\ref{active_EOM}) provides a key insight that the contribution from active disturbances is to effectively alter the strength of elongation (prefactor of $ \bten{E}^{\infty} $) and the rotation rate (prefactor of $ \bten{O}^{\infty} $).
The first prefactor is the more decisive as $ \alpha_{1} \gg |\alpha_{2}| $.
It suggests that a pusher ($ \sigma>0 $) disturbs the local shear flow in order to increase the weight of elongation. 
Once the activity exceeds the critical limit, the local shear flow transforms into an effective elongation flow whose axis of elongation points in the direction of the normalized eigenvector associated with the positive eigenvalue.
As pusher's activity further increases, the locally elongated flow asymptotically approaches the pure elongation state where the impact of $ \bten{O}^{\infty} $ is negligible in comparison (similar to our aforementioned case of $ \bten{O}^{\infty} \equiv 0 $ in a previous paragraph).
%


\subsubsection{Stability analysis of orbits}
Here we find the stability exponent of the complete non-linear Eq. (\ref{EOM}) to analyze the onset of shear-plane rotation state as shown in Fig. \ref{fig:orbits},\ref{fig:phase}.
For Newtonian fluids, $ \theta = \pi/2 $ is one of the infinite neutral Jeffery orbits.
For SOF, the polymeric stress in the fluid lifts this degeneracy. 
To quantify this, we write down the angular dynamics as
\begin{equation}
	\dot{\theta}=f(\theta,\phi), \qquad  \dot{\phi} = g(\theta,\phi), \tag{\theequation a,b}
\end{equation}
and perform a Taylor expansion near the shear-plane: $ \theta(t) = \pi/2 + \epsilon(t) $. Here $ \epsilon $ represents the deviation from Jeffery orbit and and we determine its growth \emph{i.e.,} whether it grows or shrinks with time.
At $ O(\epsilon) $ we obtain
\begin{equation}
	\frac{d \epsilon}{d t} =  \epsilon \,  \left. \frac{\partial f}{\partial \theta} \right\vert_{\theta=\pi/2},  
\end{equation}
which can be simplified and integrated over an orbit to obtain
\begin{equation} \label{eps}
	\epsilon  = \epsilon_{0} \; \exp\left[  
	\bigintssss_{0}^{-2\pi} \left( \frac{1}{g(\theta,\phi)} \frac{\partial f}{\partial \theta} \right)_{\theta=\pi/2} \, d\phi
	\right] .
\end{equation}
Here the integration is over $ -2\pi $ because the particle's rotation due to shear is in $ -\phi $ direction.
Eq.\ (\ref{eps}) can be expressed in the form of Lyapunov exponent ($ \epsilon = \epsilon_{0} \exp[\mathcal{L}  T] $) \cite{strogatz} as
\begin{equation}\label{lya}
	\mathcal{L}  \sim \frac{1}{T} \bigintsss_{0}^{-2\pi}    \left[ \frac{1}{\dot{\phi}} \frac{\partial \dot{\theta}}{\partial \theta} \right]_{\theta=\pi/2}  \;    d\phi ,
\end{equation}
where $ T $ is the time period of a Jeffery orbit $  2\pi (\lambda+\lambda^{-1})/\dot{\gamma} $.
We show the solution to Eq.\ (\ref{lya}) in the inset of Fig.\ \ref{fig:phase}.


\subsection{Kinetic model}
\subsubsection{Near-equilibrium microstructure}
First, we detail the evaluation of microstructure near  equilibrium ($ \text{Pe}_{f} = \dot{\gamma} \tau \ll 1 $) \emph{i.e.,} when stochastic reorientation dominates the shear-induced reorientation.
In this limit, we expand the orientation distribution as
\begin{equation}
	\psi = \psi_{(0)} + \text{Pe}_{f} \, \psi_{(1)} + \cdots.
\end{equation}
We substitute this in Eq. (\ref{Smol}) and collect the zeroth and first order terms in $ \text{Pe}_{f} $. At $ O(1) $, we get $ \psi_{(0)} = 1/4\pi $ as the isotropic microstructure. 
At $ O(\text{Pe}_{f}) $,  we have
\begin{equation}\label{Kinetic_model_supp}
	\nabla_{p} \cdot\Bigl\{
	\left[   \dot{\IB{p}}^{(0)} + {\rm De} \, {\rm Pe}_a \, \dot{\IB{p}}^{(1)}_{A}/8\pi  \right] \psi_{(0)}
	\Bigr\}
	- \tau D_r \nabla_{p}^{2} \psi_{(1)} + \psi_{(1)} = 0.
\end{equation}
Here we denote the equation of motion (Eq. \ref{EOM}) as having three parts: $ \dot{\IB{p}} = \dot{\IB{p}}^{(0)} + \text{De} \text{Pe}_{f} \dot{\IB{p}}^{(1)}_{P} + \text{De} \text{Pe}_{a} \dot{\IB{p}}^{(1)}_{A} $, where subscripts represent passive and active components.
Upon substituting this in Eq. (\ref{Kinetic_model_supp}), we find that the $ \dot{\IB{p}}^{(1)}_{P}  $ term only contributes at $ O(\text{Pe}_{f}^{2}) $, and is thus neglected. Upon simplification, we find that the microstructure at $ O(\text{Pe}_{f}) $ is governed by
\begin{equation}
	- \tau D_r \nabla_{p}^{2} \psi_{(1)} + \psi_{(1)} = \frac{3 }{4 \pi} (\Lambda+\alpha_{1} \text{De} \text{Pe}_{a}/8\pi) \bten{E}^{\infty} : \IB{pp} .
\end{equation}
Following \cite{chen1996rheology,nambiar2017stress}, we solve the above inhomogeneous linear differential equation using the Green's function $ G(\IB{p}|\IB{p}') $, which is governed by 
\begin{equation} \label{Green_GE}
	- \tau D_r \nabla_{p}^{2} G(\IB{p}|\IB{p}') + G(\IB{p}|\IB{p}') = \delta(\IB{p}-\IB{p}') ,
\end{equation}
where $ \delta $ represents the Dirac-delta function. 
Following \cite{arfken1999mathematical}, we expand this $ G $ into spherical harmonic series as
\begin{equation}\label{Green_expansion}
	G(\IB{p}|\IB{p}')  = \sum_{n=0}^{\infty} \sum_{m=-n}^{n} C_{n,m}(\IB{p}') Y_{n}^{m}(\IB{p}).
\end{equation}
Here $ C_{n,m} $ are the series coefficients and $ Y_{n}^{m} (\IB{p}) $ represent the spherical harmonics. 
Similarly, the Dirac-delta function can also be related to the spherical harmonics as $ \delta(\IB{p}-\IB{p}') =  \sum_{n=0}^{\infty} \sum_{m=-n}^{n} Y_{n}^{m}(\IB{p}) \overline{Y_{n}^{m}(\IB{p}')} $, where $ \overline{Y_{n}^{m}} $ represents the corresponding complex conjugate spherical harmonic \cite{arfken1999mathematical}.
Eq. (\ref{Green_GE}) is now expressible as
\begin{align}\label{Green_GE_expansion}
	\sum_{n=0}^{\infty} \sum_{m=-n}^{n} C_{n,m}(\IB{p}') \Bigl\{ -   \tau D_r & \nabla_{p}^{2} \left[ Y_{n}^{m}(\IB{p}) \right]  + Y_{n}^{m}(\IB{p})  \Bigr\} = \nonumber \\
	&  \sum_{n=0}^{\infty} \sum_{m=-n}^{n} Y_{n}^{m}(\IB{p}) \overline{Y_{n}^{m}(\IB{p}')}    	.
\end{align}
Using the properties of spherical harmonics \cite{arfken1999mathematical}, we substitute $ \nabla_{p}^{2} Y_{n}^{m} = -n(n+1) Y_{n}^{m} $, take the inner product with respect to $ \overline{Y_{N}^{M}(\IB{p})} $ on both sides of Eq. (\ref{Green_GE_expansion}),	and use the orthogonality property to obtain
\begin{align}
	C_{n,m} (\IB{p}') =  \frac{\overline{Y_{n}^{m}(\IB{p}')}}{1+ \tau D_r n(n+1)},
\end{align}
where the normalization condition yields $ C_{0,0} = 1/\sqrt{4\pi} $.
We finally use the Green's function solution to find $ \psi_{(1)} $:
\begin{equation} \label{psi_near_eq}
	\psi_{(1)}	=  \frac{3 }{4 \pi} (\Lambda+\alpha_{1} \text{De} \text{Pe}_{a}/8\pi)      \left[   \frac{   \bten{E}^{\infty}: \IB{p}\IB{p} }{1+ 6\tau D_r}  \right].
\end{equation}
In the limit of weak shear, this solution matches with the numerically obtained $ \psi $ for arbitrary $ \text{Pe}_{f} $, whose evaluation is discussed next.


\subsubsection{Numerical solution valid for arbitrary $ \text{Pe}_{f} $}
Now we detail the numerical solution of Eq.\ (\ref{Smol}), which uses the decomposition of $ \psi $ as $ \sum_{n=0}^{\infty} \sum_{m=-n}^{n} C_{n,m}(\IB{p}) Y_{n}^{m}(\IB{p}). $
We substitute this expansion in Eq. (\ref{Smol}) and use the properties of spherical harmonics to obtain
\begin{align}\label{Smol_Numerical}
	-\frac{1}{4\pi} + \sum_{n=0}^{\infty} \sum_{m=-n}^{n} C_{n,m} \left[ 1+\tau D_r n(n+1) \right] Y_{n}^{m} +& 
	\nonumber \\
	\text{Pe}_{f}  \sum_{n=0}^{\infty} \sum_{m=-n}^{n} C_{n,m}  \mathcal{H}(Y_{n}^{m}) & = 0  .
\end{align}
Here $ \mathcal{H}(Y_{n}^{m})  $ is a collection of expressions obtained by simplifying $ \nabla_{p} \cdot \left(  \dot{\IB{p}} \psi  \right)  $ . These are represented in terms of angular momentum operators for computational convenience \cite{chen1996rheology} (detailed in Supplementary Note 3).
We take an inner product with $ \overline{Y_{i}^{j}} $ on both sides of Eq. (\ref{Smol_Numerical}) and use the orthogonality property to obtain
\begin{align}
	-\frac{1}{4\pi} \int_{S} \overline{Y_{i}^{j}}  \text{d}\IB{p}
	+  C_{i,j} \left[ 1+\tau D_r i(i+1) \right] +& 
	\nonumber \\
	\text{Pe}_{f}  \sum_{n=0}^{\infty} \sum_{m=-n}^{n} C_{n,m} \int_{S} \overline{Y_{i}^{j}}  \mathcal{H}(Y_{n}^{m})  \text{d}\IB{p}  &  =0.
\end{align}
Since the $ n=0 $ mode is already known from normalization, the above equations can be recasted as the following linear system of equations where $ C_{i,j} $ (for $ i \geq 1 $) is unknown:
\begin{align}
	C_{i,j} \left[ 1+\tau D_r i(i+1) \right] + \; \; & 
	\nonumber \\
	\text{Pe}_{f}  \sum_{n=1}^{100} \sum_{m=-n}^{n} C_{n,m} \int_{S} \overline{Y_{i}^{j}} & \mathcal{H}(Y_{n}^{m})  \text{d}\IB{p}    =
	\nonumber \\
	-	& \text{Pe}_{f}  C_{0,0} \int_{S} \overline{Y_{i}^{j}}  \mathcal{H}(Y_{0}^{0})  \text{d}\IB{p}  .
	,
\end{align}
To find $ C_{i,j} $ coefficients, the above equations are solved using the Clebsch-Gordon coefficient formulation, where first 100 modes in $ n $ are used.
We note that although the numerical approach is valid for arbitrary $ \text{Pe}_{f} $, there is an indirect upper bound of weak viscoelasticity 
that we must adhere to, as we are using the SOF model. 
In non-dimensional terms this bound is: $ \text{De} \, \text{Pe}_{f}  < 1 $.
Thus, for $ \text{De}=0.1 $, we explore the results within an upper bound of $ \text{Pe}_f = 10 $.


\subsection{Rheology}
First we show the expanded version of the flow-induced component of the particle stress 
\begin{equation}
	\bten{\Sigma}^{F}_{p} = -\mu_f n l^{3} A \dot{\gamma}    {\overset{\nabla}{\bten{a}_{2}  }} /2 
\end{equation} where $ \bten{a}_{i} $ is a shorthand notation for the ensemble of orientation moment of $ i^{\text{th}} $ order: $ \ang{\IB{p}^{\otimes i}} $. Expanding the upper-convected derivative, we obtain
\begin{align} \label{GF}
	\!   {\overset{\nabla}{\bten{a}_2 }}  
	= &  \left(\Lambda - 1 \right)  \left(  \bten{E}^{\infty} \cdot \bten{a}_{2}  + \bten{a}_{2} \cdot  \bten{E}^{\infty}  \right)  -  2 \Lambda \bten{a}_{4} : \bten{E}^{\infty}
	\nonumber \\
	& \!\!\! \!\!\!\!\!\!    + \text{De} \Big[ \text{Pe}_{f} \beta_{1}  \left(  \bten{E}^{\infty} \cdot \bten{a}_{4} : \bten{E}^{\infty}  - 2 \bten{E}^{\infty} : \bten{a}_{6} : \bten{E}^{\infty} +   \bten{E}^{\infty} :  \bten{a}_{4} \cdot \bten{E}^{\infty}  \right)  	\nonumber \\
	& \!\!\! \!\!\!\!\!\!  
	+ \text{Pe}_{f} \beta_{2} \left( \bten{O}^{\infty} \cdot \bten{a}_{4} : \bten{E}^{\infty}  -  \bten{E}^{\infty} : \bten{a}_{4} \cdot \bten{O}^{\infty}   \right)
	\nonumber \\
	& \!\!\! \!\!\!\!\!\!    + \text{Pe}_{a} \alpha_{1}  \left(  \bten{E}^{\infty} \cdot \bten{a}_{2}   -  2 \bten{a}_{4}:\bten{E}^{\infty}   + \bten{a}_{2} \cdot \bten{E}^{\infty}    \right)
	\nonumber \\
	& \!\!\! \!\!\!\!\!\!    + \text{Pe}_{a} \alpha_{2}  \left(  \bten{O}^{\infty} \cdot \bten{a}_{2}   -  \bten{a}_{2} \cdot \bten{O}^{\infty}
	\right)
	\Big].
\end{align}
This $ \bten{\Sigma}^{F}_{p}  $ matches with the Newtonian bulk stress for $ \text{De}=0 $ \cite{hinch1976constitutive}.
For near-equilibrium results ($ \text{Pe}_{f} \ll 1 $), the above expression is substituted in Eq. (\ref{visc_res}) and orientation distribution from Eq. $ (\ref{psi_near_eq}) $ is used to perform the ensemble integral.
Since the passive viscoelastic contribution is $ O(\text{Pe}_{f}^{2}) $, it does not contribute at the $ O(\text{Pe}_{f}) $ calculations.
The viscosity relates to the deviatoric particle-induced stress as $ 	\varphi\mu_{p} = {\Sigma_{p \, xy}} /{\dot{\gamma}}  , $ where $ \varphi=n  a^{3} $ is the volume fraction.
This viscometeric relation can be simplified to obtain the near-equilibrium viscosity ratio
	\begin{align}
		\frac{\mu_{p}}{  \mu_{f} } =  	&\frac{ 4A(5-3\Lambda) }{15} +   \frac{  48 A \tau D_{r} - \text{Pe}_{a} / \pi}{5\left( 1+ 6 \tau D_r \right)}  \left( \Lambda + \frac{\text{De} \, \text{Pe}_{a} \alpha_{1}}{8\pi}  \right)  \nonumber \\
		& -  \frac{\text{De} \, \text{Pe}_{a} A \alpha_{1}}{10\pi}. 
	\end{align}
	The last term in the above expression is negligible and is generated from the non-Newtonian component of $ \bten{\Sigma}_{p}^{F} $. 
	Thus, the major modifications due to fluid's viscoelasticity scale as $ \text{Pe}_{a}^{2} $, as also depicted in Fig. \ref{fig: viscosity}d.
	
	\medskip
	
	\section*{Acknowledgements}
	Support from the Alexander von Humboldt Foundation is gratefully acknowledged. We also acknowledge financial support from the Collaborative Research Center 910 funded by the Deutsche Forschungsgemeinschaft.
	S.N acknowledges support from the Swedish Research Council (under grant no. 638-2013-9243) and Nordita, which is partially supported by NordForsk.
	A.C thanks Jonas Einarsson for his advice on the effective equation of motion.


		\clearpage

\pagebreak
\widetext
\begin{center}
	\textbf{\large {Supplementary material for: ``Orientational dynamics and rheology of  active suspensions in weakly viscoelastic flows''}}
\end{center}
\setcounter{equation}{0}
\setcounter{figure}{0}
\setcounter{table}{0}
\setcounter{page}{1}
\makeatletter
\renewcommand{\theequation}{S\arabic{equation}}
\renewcommand{\thefigure}{S\arabic{figure}}

\section{Coefficients of the multipole expansion}
The tensorial coefficients of the spheroidal multipoles are
\begin{subequations}
	\begin{equation}
		\mathsf{p}^{a}_{jklm} = (p_{j} p_{k} - \delta_{jk}/3) (p_{l} p_{m} - \delta_{lm}/3),
	\end{equation}
	\begin{equation}
		\mathsf{p}^{b}_{jklm} = p_{j} p_{l} \delta_{km} + p_{j} p_{m} \delta_{kl} + p_{k} p_{l} \delta_{jm} + p_{k} p_{m} \delta_{jl} - 4p_{j} p_{k} p_{l} p_{m},
	\end{equation}
	\begin{equation}
		\mathsf{p}^{c}_{jklm} = -\delta_{jk} \delta_{lm} + \delta_{jl} \delta_{km} + \delta_{jm} \delta_{kl} + p_{l} p_{m} \delta_{jk} + p_{j} p_{k} \delta_{lm} - p_{j} p_{m} \delta_{kl} - p_{k} p_{m} \delta_{jl} - p_{j} p_{l} \delta_{km} - p_{k} p_{l} \delta_{jm} + p_{j} p_{k} p_{l} p_{m}.
	\end{equation}
\end{subequations}
The other scalar coefficients are
\begin{subequations}
	\begin{equation}
		\mathbb{A}^{R} = \frac{\sqrt{\lambda^{2}-1}}{4(C-\lambda^{3}+\lambda)},
		\;\; \mathbb{B}^{R} = \frac{\sqrt{\lambda^{2}-1}(\lambda^{2}+1)}{4(C+\lambda^{3}-\lambda-2C\lambda^{2})},
		\;\; \mathbb{C}^{R} = \frac{(\lambda^{2}-1)^{3/2}}{4(C+\lambda^{3}-\lambda-2C\lambda^{2})},
	\end{equation}
	\begin{equation}
		\mathbb{A}^{S} = \frac{(\lambda^{2}-1)^{3/2}}{4(C-3\lambda^{3}+3\lambda+2C\lambda^{2})},
		\;\; \mathbb{B}^{S} = \frac{-(\lambda^{2}-1)^{3/2} (C\lambda+\lambda^{4}-3\lambda^{2}+2)}{8(C+\lambda^{3}-\lambda-2C\lambda^{2})(-3C\lambda + \lambda^{4} + \lambda^{2}-2)},
		\;\; \mathbb{C}^{S} = \frac{(\lambda^{2}-1)^{3/2}}{2(3C +2\lambda^{5}-7\lambda^{3}+5\lambda)}.
	\end{equation}	
\end{subequations}
Here $ C=\sqrt{\lambda^{2}-1} \coth^{-1}\left[ \lambda/\sqrt{\lambda^{2}-1} \right] $.

\section{Further results of orientation dynamics}

\subsection{Coefficients of equation of motion}

\begin{figure}[h]
	\centering
	\includegraphics[width=0.7\textwidth]{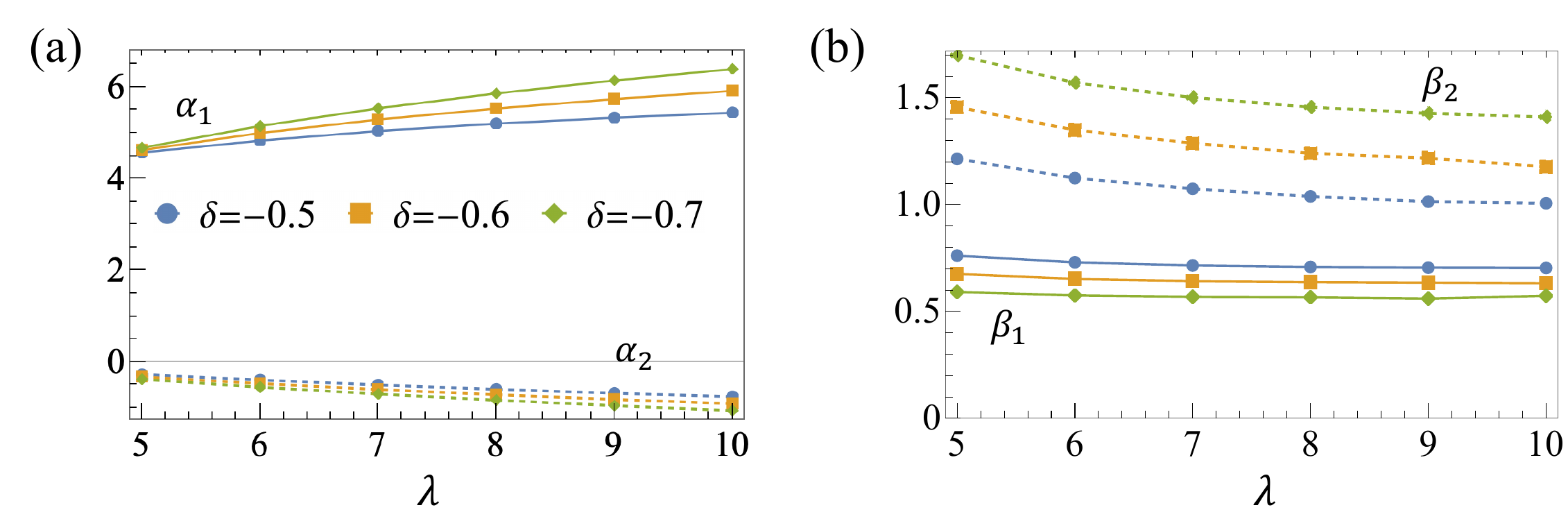}
	\caption{Coefficients of equation of motion (Eq. 1 in main text). $ \alpha_{i} $ represent active coefficients and $ \beta_{i} $ are passive coefficients.}	
	\label{fig:supp_coeff}
\end{figure}

Supplementary Figure \ref{fig:supp_coeff} shows the coefficients of the analytical equation of motion. They vary weakly with aspect ratio $ \lambda $ and the viscometric parameter $ \delta $, which typically lies between $ -0.7 $ and $ -0.5 $.
Since the coefficients vary weakly with  $ \lambda $ and the integration time is very large for $  \lambda >10 $, we consider aspect ratios between 5 and 10.

\subsection{Alignment angle}

\begin{figure}[h]
	\centering
	\includegraphics[width=0.4\textwidth]{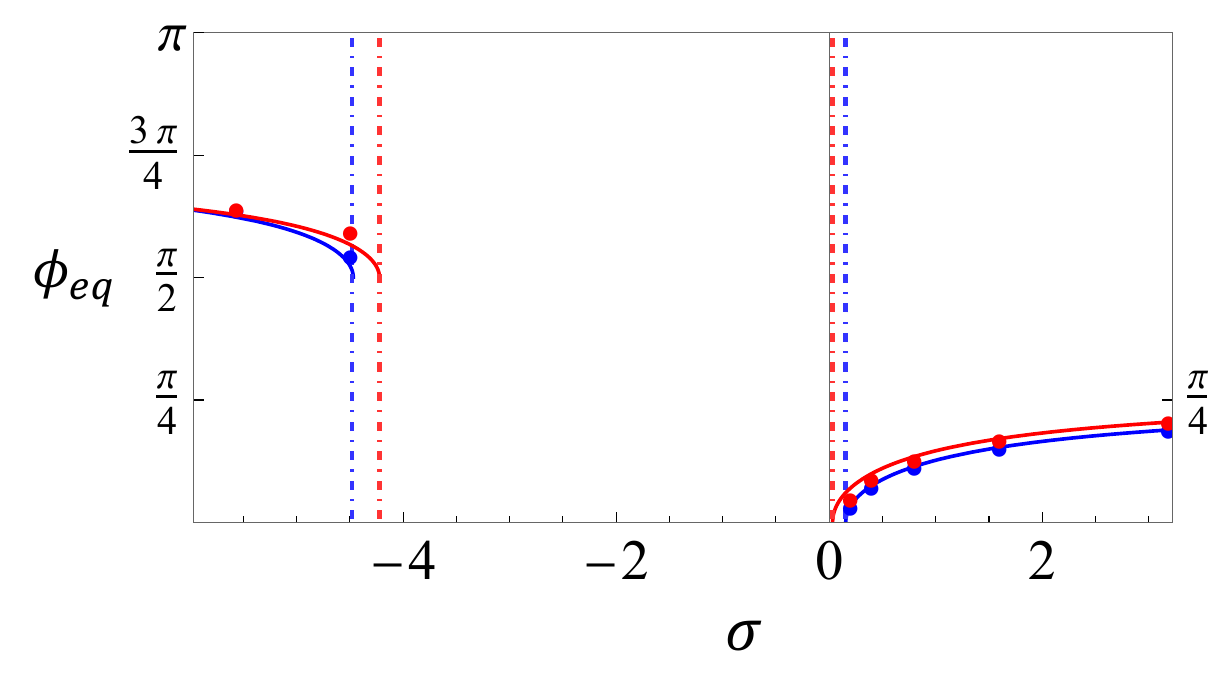}
	\caption{Comparison of alignment angles obtained by integrating the full equation of motion (dots) with those found using analytical approximation (line). Blue: $ \lambda=5 $, red: $ \lambda=10 $. Vertical dashed lines separate the regions of alignment. 
		Other parameters: $\delta=-0.6$, $ \text{Wi}=0.1 $ .}
	\label{fig:supp_phase}
\end{figure}
Supplementary Figure \ref{fig:supp_phase} shows the comparison of angles of alignment obtained by integrating the analytical equation of motion (Eq. \textcolor{red}{1} of main text) and those from the eigenvalue analysis. 
Note that the phase diagram has a weak dependency on aspect ratio.

\subsection{Passive spheroid in weakly viscoelastic shear flow}

The non-dimensional equation of motion for the orbits is 
\begin{equation}
	\dot{\IB{p}} =  \left(  \bten{I} -  \IB{pp}  \right) \cdot  \left(   \bten{E}^{\infty} \cdot \IB{p}   \right) \left[  \Lambda + \text{Wi}  \, \sigma  \alpha_{1} + \text{Wi} \, \beta_1 \,  \bten{E}^{\infty}: \IB{pp} \right] 
	+ \IB{\Omega}^{\infty} \times \IB{p} \left[  1 + \text{Wi} \, \sigma  \alpha_{2} + \text{Wi} \, \beta_2 \,  \bten{E}^{\infty}: \IB{pp}   \right].
\end{equation}
This can be written in the trigonometric form as
\begin{subequations}
	\begin{equation}
		\dot{\theta}	=  \frac{\Lambda}{4} \sin 2 \theta  \sin 2 \phi  +   \text{Wi} \,\beta_{1} \sin ^3\theta \cos \theta \sin ^2\phi \cos ^2\phi
	\end{equation}
	\begin{equation}
		\dot{\phi}	=   -\frac{1 }{\lambda^2+1} \left[  \lambda^2 \sin ^2\phi+\cos ^2\phi\right]  + \frac{\text{Wi}}{4}  \left[ \beta_{1} \cos (2 \phi )- \beta_{2} \right] \sin ^2\theta  \sin 2 \phi  .
	\end{equation}
\end{subequations}
To quantify the log-rolling and compare it with Brunn \cite{brunn1977slow}, we use the following transformation \cite{leal1971effect}:
\begin{equation}
	\theta = \arctan \left( \frac{C \lambda}{\sqrt{\lambda^{2} \cos^{2}\phi + \sin^{2} \phi}} \right), \quad  \tau = \arctan \left( \frac{\tan \phi}{\lambda} \right)   .
\end{equation}
Here $ C $ represents the orbit constant ($ C=0 $ and $ C=\infty $ being the log-rolling and shear-plane rotation orbits, respectively) and $ \tau $ is essentially the time coordinate that is scaled with $ {T}/{2\pi}  $, where $ T $ is the Jeffery orbit.
Using the above relation, we obtain the evolution equation of the orbit constant as
\begin{align} \label{Ctau}
	\dot{C}(t) &= \text{Wi} \frac{ \left[C^3 \left(\beta_{1} \lambda^2-\beta_{2}\lambda^2+\beta_{1}+\beta_{2}\right) \sin ^2\tau \cos ^2\tau\right]}{2 \left(\lambda^2 C^2 \cos ^2\tau+C^2 \sin ^2\tau+1\right)}
	, \nonumber \\
	\dot{\tau}(t) &= \frac{\lambda}{1+\lambda^{2}}  +  \text{Wi} \left[ \frac{ \sin \tau\cos \tau\left(2 \beta_{1} \lambda^{2} C^2 \cos ^2\tau+\beta_{1}+\beta_{2}\right)}{2 C^2 \left(\lambda^{2} \cos ^2\tau+\sin ^2\tau\right)+2} -
	\frac{1}{4}  (\beta_{1}+\beta_{2}) \sin 2 \tau \right] .
\end{align}
We now compare our result with Brunn \cite{brunn1977slow} who, for a rigid tri-dumbbell, showed that the particle slowly drifts towards the log-rolling orbit.
We plot Eq. (\ref{Ctau}) and Eq. (4.17) from Brunn in Supplementary Figure \ref{fig: Brunn} and find a good match. 
Note that over one period the particle's orbit declines in a wiggly manner. 
Interestingly, these wiggles have a physical significance and can be characterized as orbit reduction with two distinct plateaus. 
Supplementary Figure \ref{fig: Brunn}b shows that the long  plateau (around $ t/T = 1/4 $) occurs first and is the state where particle is in the vicinity of flow-vorticity plane (where it slows down and spends most of its time).
Once the particle exits the alignment state, it flips quickly to the other side ($ -x $ axis). This flip generates the short plateau around $ t/T = 1/2 $.

\begin{figure}[h]
	\centering
	\includegraphics[width=0.67\textwidth]{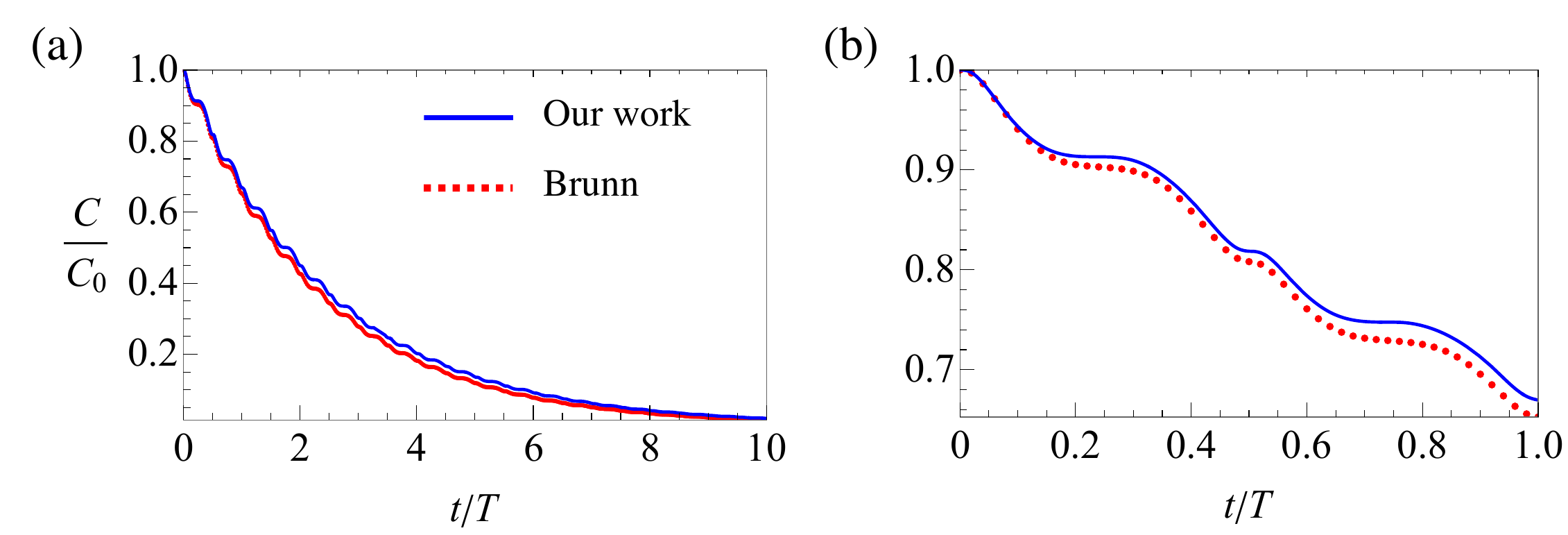}
	\caption{Comparison with Ref. \cite{brunn1977slow} for the variation of the orbit constant $ C/C_{0} $, where $ C_{0} $ is the orbit constant for $ t=0 $. (a) Variation over 10 periods. (b) Variation over one period.
		Other parameters: $\lambda=5$, $ T= 2\pi \frac{\lambda^{2}+1}{\lambda} $, $\delta=-0.6$, $ \text{Wi}=0.1 $.}
	\label{fig: Brunn}
\end{figure}

.

\section{Details of Kinetic model}
Here we provide full expressions and further details involved in the computational steps of kinetic model.	
The spectral decomposition of $ \psi $ is carried out as
\begin{equation}
	\psi(\theta,\phi) =  \sum_{n=0}^{\infty} \sum_{m=-n}^{n} C_{n,m}(\IB{p}) Y_{n}^{m}(\IB{p}).
\end{equation}
Substituting this in the Smoluchowski equation of main text and using the properties of spherical harmonics yields
\begin{align}\label{Smol_Numerical}
	-\frac{1}{4\pi} + \sum_{n=0}^{\infty} \sum_{m=-n}^{n} C_{n,m} \left[ 1+\tau D_r n(n+1) \right] Y_{n}^{m} +
	\text{Pe}_{f}  \sum_{n=0}^{\infty} \sum_{m=-n}^{n} C_{n,m}  \mathcal{H}(Y_{n}^{m})  = 0  .
\end{align}
Here $ \mathcal{H}(Y_{n}^{m})  $ is a term obtained by simplifying $ \nabla_{p} \cdot \left(  \dot{\IB{p}} \psi  \right)  $ and is represented in terms of angular momentum operators  for computational convenience \cite{chen1996rheology}. 
\begin{align}\label{Hfunc}
	\mathcal{H}(Y_{n}^{m}) & =	 \left[ -\frac{\text{i} \, \mathfrak{L}_{z}}{2} + \frac{ \text{i}\Lambda}{2}  \left(3\text{i} {Y_{n}^{m}}  \sin ^2(\theta ) \sin (2 \phi )+\mathfrak{L}_{y} \sin (2 \theta ) \sin (\phi )+\mathfrak{L}_{z} \cos (2 \theta ) \sin ^2(\phi )+\mathfrak{L}_{z} \cos ^2(\phi )\right)  \right] + 
	\nonumber \\
	& \!\!\!\!\!\!\!\!\!\!\!\!\!\!   \text{De} \Bigg\lbrace \frac{\text{i} \, \text{Pe}_{a} }{16\pi} \left[  \alpha_{1} \left( 3\text{i}Y_{n}^{m}  \sin ^2(\theta ) \sin (2 \phi )+\mathfrak{L}_{y} \sin (2 \theta ) \sin (\phi )+\mathfrak{L}_{z} \cos (2 \theta ) \sin ^2(\phi )+\mathfrak{L}_{z} \cos ^2(\phi ) \right)  -\alpha_2 \mathfrak{L}_{z} \right] + 
	\nonumber \\
	& \!\!\!\!\!\!\!  \frac{\text{i} \, \text{Pe}_{f}}{16}  \Big[
	\beta_{1} \sin ^2(\theta ) \Big(-i Y_{n}^{m}  (3 + 5 \cos (2 \theta ))-10\text{i}Y_{n}^{m}  \sin ^2(\theta ) \cos (4 \phi )+2 \mathfrak{L}_{y} \sin (2 \theta ) \cos (\phi )-2 \mathfrak{L}_{y} \sin (2 \theta ) \cos (3 \phi )+
	\nonumber \\
	&   \quad
	2 \mathfrak{L}_{z} \sin ^2(\theta ) \sin (4 \phi )  +2 \mathfrak{L}_{z} \cos (2 \theta ) \sin (2 \phi )+2 \mathfrak{L}_{z} \sin (2 \phi )\Big)
	+ \beta_{2} \sin ^2(\theta ) \left(8\text{i}Y_{n}^{m}  \cos (2 \phi ) -4 \mathfrak{L}_{z} \sin (2 \phi ) \right)
	\Big]
	\Bigg\rbrace .
\end{align}
Here, the angular momentum operators ($ \mathfrak{L} $) are shorthand notation for the following gradients on harmonic $ Y_{n}^{m} $:
\begin{align}
	\mathfrak{L}_{y} =  L_{y}|Y_{n}^{m}\rangle = -\text{i} \cos \phi \frac{\partial Y_{n}^{m}}{\partial \theta} + \text{i} \cot \theta \sin \phi \frac{\partial Y_{n}^{m}}{\partial \phi} \quad \mbox{and\ } \quad \mathfrak{L}_{z}=  L_{z}|Y_{n}^{m}\rangle  = - \text{i} \frac{\partial Y_{n}^{m}}{\partial \phi}.
\end{align}
The trigonometric functions in the Eq. ({\ref{Hfunc}}) are converted into spherical harmonics. For example:
\begin{equation}
	\sin \theta \cos \theta \sin \phi = \text{i} \sqrt{\frac{2\pi}{15}} (Y_{2}^{-1}+Y_{2}^{1}), \quad \cos^{2} \theta = \frac{1}{3} \left( 2 \sqrt{4\pi/5} Y_{2}^{0} + \sqrt{4\pi} Y_{0}^{0} \right).
\end{equation}
Using such transformations, Eq. (\ref{Hfunc}) yields
{\begin{align}\label{Hfunc_harmonic}
		\mathcal{H}(Y_{n}^{m}) & =  \Bigg[ -i \sqrt{\pi } \mathfrak{L}_{z} Y^{0}_{0} + \Lambda \Bigg( -\sqrt{\frac{2 \pi }{15}} \mathfrak{L}_{y} Y_{2}^{-1}-\sqrt{\frac{2 \pi }{15}} \mathfrak{L}_{y} Y_{2}^{1}+i \sqrt{\frac{2 \pi }{15}} Y_{2}^{-2} (\mathfrak{L}_{z}-3 Y_{n}^{m} )+i \sqrt{\frac{2 \pi }{15}} Y_{2}^{2} (\mathfrak{L}_{z}+3 Y_{n}^{m} )
		\nonumber \\
		& \qquad \qquad \qquad \qquad \quad +\frac{1}{3} i \sqrt{\pi } \mathfrak{L}_{z} Y_{0}^{0}+\frac{2}{3} i \sqrt{\frac{\pi }{5}} \mathfrak{L}_{z} Y_{2}^{0} \Bigg) \Bigg] \; + 
		\nonumber \\
		&   \text{De}  \Bigg\lbrace  \text{Pe}_{a}  \Bigg[  -\frac{\text{$\alpha $1} \mathfrak{L}_{y} Y_{2}^{-1}}{4 \sqrt{30 \pi }}-\frac{\text{$\alpha $1} \mathfrak{L}_{y} Y_{2}^{1}}{4 \sqrt{30 \pi }}+\frac{i \mathfrak{L}_{z} Y_{0}^{0} (\text{$\alpha $1}-3 \text{$\alpha $2})}{24 \sqrt{\pi }} +\frac{i \text{$\alpha $1} \mathfrak{L}_{z} Y_{2}^{-2}}{4 \sqrt{30 \pi }}+\frac{i \text{$\alpha $1} \mathfrak{L}_{z} Y_{2}^{0}}{12 \sqrt{5 \pi }}  +\frac{i \text{$\alpha $1} \mathfrak{L}_{z} Y_{2}^{2}}{4 \sqrt{30 \pi }} + 
		\nonumber \\
		&  \qquad\qquad -\frac{1}{4} i \sqrt{\frac{3}{10 \pi }} \text{$\alpha $1} Y_{n}^{m}  Y_{2}^{-2}+ \frac{1}{4} i \sqrt{\frac{3}{10 \pi }} \text{$\alpha $1} Y_{n}^{m}  Y_{2}^{2}
		\Bigg] \; + 
		\nonumber \\
		&  \qquad \;  \text{Pe}_{f} \Bigg[ \frac{1}{7} i \sqrt{\frac{2 \pi }{15}} \beta_{1} \mathfrak{L}_{y} Y_{2}^{-1}-\frac{1}{7} i \sqrt{\frac{2 \pi }{15}} \beta_{1} \mathfrak{L}_{y} Y_{2}^{1}-\frac{1}{3} i \sqrt{\frac{\pi }{35}} \beta_{1} \mathfrak{L}_{y} Y_{4}^{-3}-\frac{1}{21} i \sqrt{\frac{\pi }{5}} \beta_{1} \mathfrak{L}_{y} Y_{4}^{-1}+ \frac{1}{21} i \sqrt{\frac{\pi }{5}} \beta_{1} \mathfrak{L}_{y} Y_{4}^{1}+
		\nonumber \\
		& \qquad \qquad \;    \frac{1}{3} i \sqrt{\frac{\pi }{35}} \beta_{1} \mathfrak{L}_{y} Y_{4}^{3}-\frac{1}{7} \sqrt{\frac{\pi }{30}} \mathfrak{L}_{z} Y_{2}^{-2} (\beta_{1}-7 \beta_{2})+\frac{1}{7} \sqrt{\frac{\pi }{30}} \mathfrak{L}_{z} Y_{2}^{2} (\beta_{1}-7 \beta_{2})  -\frac{1}{3} \sqrt{\frac{2 \pi }{35}} \beta_{1} Y_{4}^{-4} (\mathfrak{L}_{z}-5 Y_{n}^{m} )+
		\nonumber \\
		&  \qquad\quad \;
		\frac{1}{3} \sqrt{\frac{2 \pi }{35}} \beta_{1} Y_{4}^{4} (\mathfrak{L}_{z}+5 Y_{n}^{m} )-\frac{1}{21} \sqrt{\frac{2 \pi }{5}} \beta_{1} \mathfrak{L}_{z} Y_{4}^{-2}+\frac{1}{21} \sqrt{\frac{2 \pi }{5}} \beta_{1} \mathfrak{L}_{z} Y_{4}^{2}+\frac{2}{7} \sqrt{\frac{\pi }{5}} \beta_{1} Y_{n}^{m}  Y_{2}^{0} -\frac{2}{21} \sqrt{\pi } \beta_{1} Y_{n}^{m}  Y_{4}^{0} +
		\nonumber \\
		&  \qquad \quad\;
		-\sqrt{\frac{2 \pi }{15}} \beta_{2} Y_{n}^{m}  Y_{2}^{-2}-\sqrt{\frac{2 \pi }{15}} \beta_{2} Y_{n}^{m}  Y_{2}^{2}
		\Bigg]
		\Bigg\rbrace
\end{align}}
The angular momentum operators are further simplified using the following relations \cite{arfken1999mathematical}
\begin{align}
	\mathfrak{L}_{y} =  L_{y}|Y_{n}^{m}\rangle = \frac{1}{2} \sqrt{n(n+1)-m(m+1)} Y_{n}^{m+1} - \frac{1}{2} \sqrt{n(n+1)-m(m-1)} Y_{n}^{m-1}
	\quad \mbox{and\ } \quad \mathfrak{L}_{z}=  L_{z}|Y_{n}^{m}\rangle  =  m Y_{n}^{m}.
\end{align}
We substitute above relations in Eq. (\ref{Hfunc_harmonic}) to simplify $ \mathcal{H}(Y_{n}^{m}) $ and use that in Eq. (\ref{Smol_Numerical}).
Next, we take an inner product with $ \overline{Y_{i}^{j}} $ on both sides of Eq. (\ref{Smol_Numerical}) and use the orthogonality property to obtain
\begin{align}
	-\frac{1}{4\pi} \int_{S} \overline{Y_{i}^{j}}  \text{d}\IB{p}
	+  C_{i,j} \left[ 1+\tau D_r i(i+1) \right] +
	\text{Pe}_{f}  \sum_{n=0}^{\infty} \sum_{m=-n}^{n} C_{n,m} \int_{S} \overline{Y_{i}^{j}}  \mathcal{H}(Y_{n}^{m})  \text{d}\IB{p}   =0.
\end{align} 
The first term is only non-zero for the zeroth mode ($ i=0 $), which yields $ -1/\sqrt{4\pi} $.
Since the $ C_{0,0}=1/\sqrt{4\pi} $ is already known from normalization, the first term cancels the $ i=0 $ mode of second term.
Using these identities, Eq. (14) can be recasted as the following linear system of equations in terms of $ C_{i,j} $:
\begin{align}
	C_{i,j} \left[ 1+\tau D_r i(i+1) \right] + \; \; 
	\text{Pe}_{f}  \sum_{n=1}^{100} \sum_{m=-n}^{n} C_{n,m} \int_{S} \overline{Y_{i}^{j}}  \mathcal{H}(Y_{n}^{m})  \text{d}\IB{p}    =      -   \text{Pe}_{f}  C_{0,0} \int_{S} \overline{Y_{i}^{j}}  \mathcal{H}(Y_{0}^{0}) \text{d}\IB{p},
	\quad   \mbox{ for\ }  i \geq 1,
\end{align}
where the right hand side is known.
The two integral terms in the above set of equations contain the inner product of three spherical harmonics. These are represented in terms of Clebsch-Gordon coefficients that exploit the orthogonal relations between harmonics for computational ease \cite{arfken1999mathematical}:
\begin{equation}
	\int_{0}^{\pi} \int_{0}^{2\pi}  Y_{N}^{M} Y_{a}^{b} Y_{n}^{m} \, \sin \theta \, \text{d}\phi \text{d}\theta = \sqrt{\frac{(2 a+1) (2 n+1)}{4 \pi  (2 \text{N}+1)}} 
	\begin{bmatrix}
		a & n & N \\
		0 & 0 & 0
	\end{bmatrix}
	\begin{bmatrix}
		a & n & N \\
		b & m & M
	\end{bmatrix}
	.
\end{equation}
The bracket terms here are the Clebsch-Gordan coefficients inbuilt in Mathematica 13.

\bigskip

Below we show the match between our results (obtained from the above kinetic model) in the limit $ De=0 $ and those obtained by Saintillan \cite[Fig. 5a, $ \tilde{\tau}=1 $]{saintillan2010dilute} .

\begin{figure}[h]
	\centering
	\includegraphics[width=0.3\textwidth]{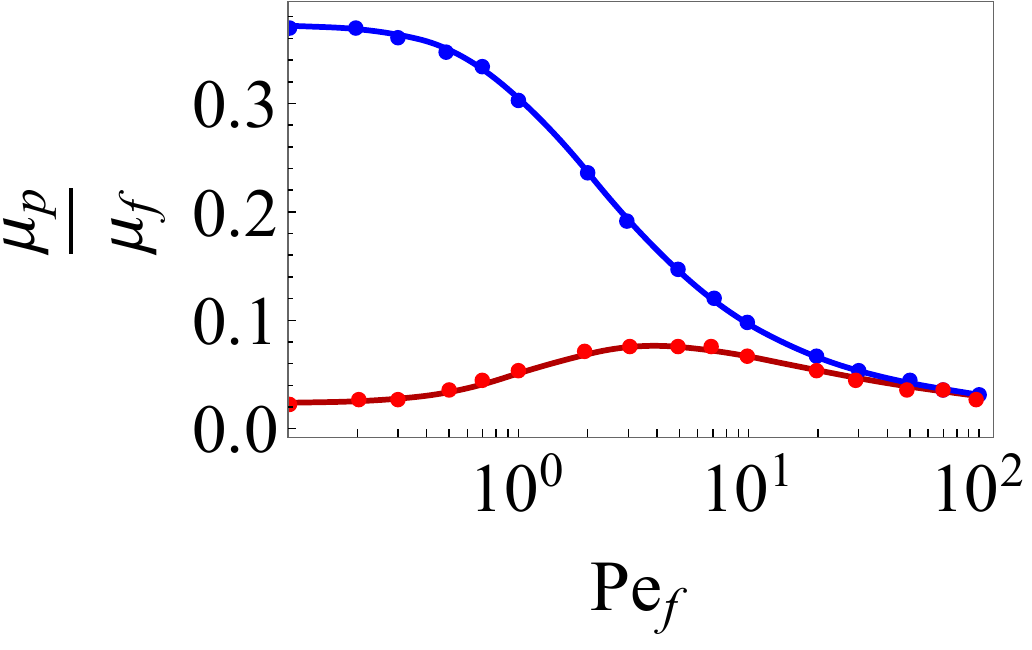}
	\caption{Comparing the shear viscosity as a function of $ \text{Pe}_f $ from our analysis in the Newtonian fluid limit with that of Ref. \cite{saintillan2010dilute} for $ \frac{|\text{Pe}_{a}|}{8\pi A}= 1 $, $\Lambda= 1 $, $\tau=1$, $ D_{r}=0.1 $. Lines (blue is puller, red is pusher) indicate our viscosity ratio and dots correspond to Ref. \cite{saintillan2010dilute}, whose viscosity is defined here with respect to the particle volume fraction $ \varphi= n a^{3} $. Ref. \cite{saintillan2010dilute} used $ \varphi = n l^{3} A $.}
	\label{fig: viscosity}
\end{figure}

\section{Log-rolling state in the microstructure}

\begin{figure}[h]
	\centering
	\includegraphics[width=0.9\textwidth]{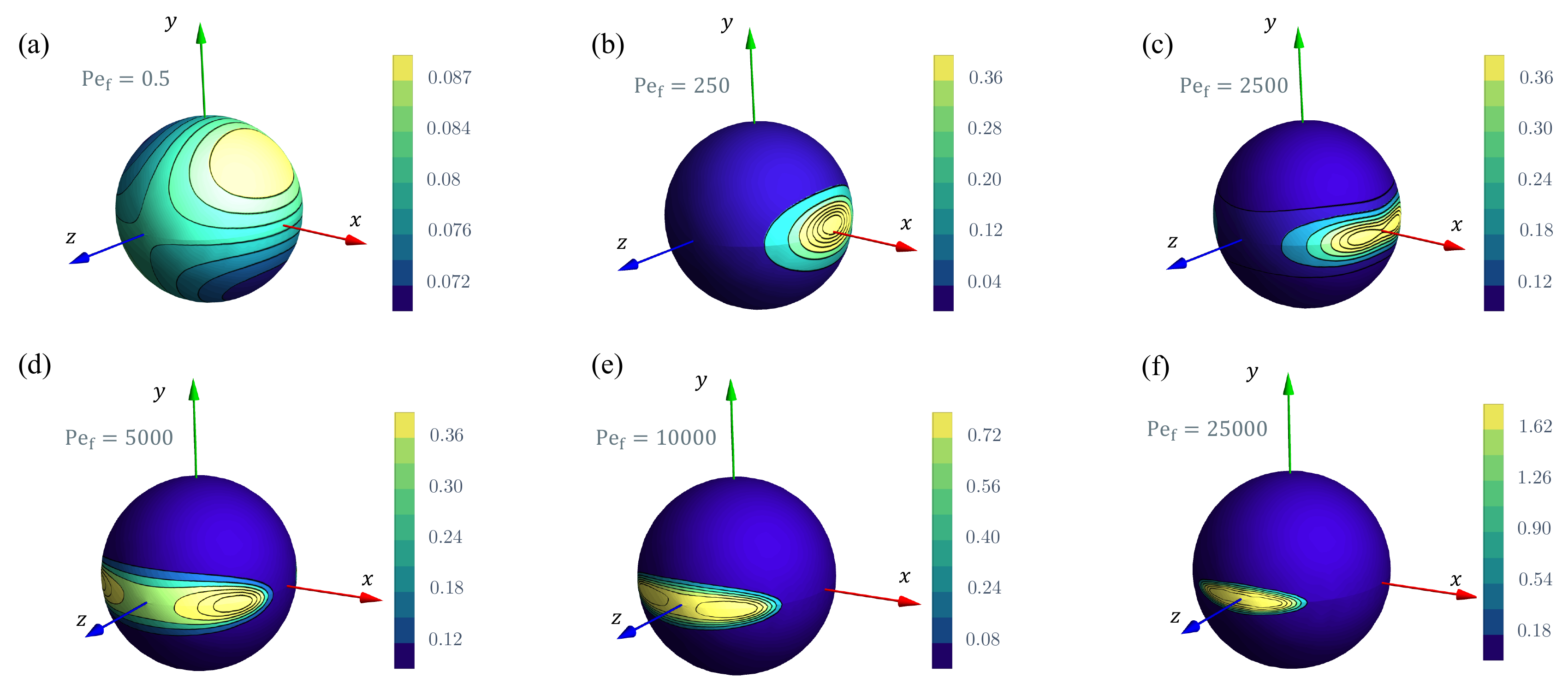}
	\caption{Orientation distribution $\psi(\IB{p})$ of a passive particle with increasing shear rate ($ \text{Pe}_{f} $) for $ \text{De}= 2 \times 10^{-5} $, $\lambda=5$, $ D_{r}=2 \times 10^{-4} \, $s$ {}^{-1}  $ ($ l = 30 \mu $m).}
	\label{fig:LR}
\end{figure}

In Supplementary Figure \ref{fig:LR}, we illustrate how log-rolling state manifests in the microstructure for large (weakly Brownian) passive particles. We consider a larger particle of length $ 30 \mu $m, which corresponds to $ D_{r}=2 \times 10^{-4} \, $s$ {}^{-1} $.  As discussed in the main text, Supplementary Figure \ref{fig:LR}a shows that when shear is weak, the isotropic distribution of rods is modified primarily by extensional component ($ \bten{E}^{\infty} $) of the shear flow. Hence, the microstructure peaks near the extensional axis. 
As the shear rate increases, the rotational flow ($ \IB{\Omega}^{\infty} $) contributes 
and the peak shifts towards the flow axis. 
Supplementary Figure \ref{fig:LR}b shows that particle almost aligns with the flow direction (positive and negative x-axis). Further 
increasing $ \text{Pe}_{f}$ spreads the distribution in the flow-vorticity plane (Supplementary Figure \ref{fig:LR}c) and, ultimately for $ \text{Pe}_{f} > 10^{3} $,
a peak develops along the vorticity axis, which is entirely reminiscent of log rolling (Supplementary Figure \ref{fig:LR}d,e). A further increase concentrates the orientation distribution near the vorticity axis (Supplementary Figure \ref{fig:LR}f).

\newpage



\end{document}